\documentclass[lettersize,journal]{IEEEtran}
\usepackage{amsmath,amsfonts}
\usepackage{amsthm}
\usepackage{algorithm}
\usepackage[noend]{algpseudocode}
\usepackage{array}
\usepackage[caption=false,font=normalsize,labelfont=sf,textfont=sf]{subfig}
\usepackage{textcomp}
\usepackage{stfloats}
\usepackage{url}
\usepackage{verbatim}
\usepackage{tikz}
\usepackage{booktabs}
\usepackage{graphicx}
\usepackage{cite}
\newtheorem{theorem}{\textbf{Theorem}}

\newtheorem{remark}{\textbf{Remark}}
\newtheorem{definition}{\textbf{Definition}}
\newtheorem{assumption}{Assumption}

\begin{document}

\title{Operational Reliability of Deadline-Constrained Task Assignment: Stability Characterization and Adversarial Routing}

\author{Roee M. Francos*, Daniel Garces*, Orhan Eren Akgün and Stephanie Gil
\thanks{This work is partially supported by AFOSR award $\#$FA9550-22-1-0223 and DARPA YFA award $\#$D24AP00319-00.}

\thanks{(*Co-primary authors) R.M.~Francos, D.~Garces, O.~Akgün, and S.~Gil are with the School of Engineering and Applied Sciences, Harvard University, Cambridge, MA 02138 USA (e-mails: {\tt\small rfrancos@seas.harvard.edu, dgarces@g.harvard.edu, orhan.eakgun@gmail.com, sgil@seas.harvard.edu}).}
}



\maketitle

\begin{abstract}
Automated task-assignment systems often serve stochastic tasks subject to finite deadlines. In these settings, conventional backlog-based stability can be misleading: finite task lifetimes may keep the number of outstanding tasks bounded even as deadline failures continue indefinitely, while average failure-rate criteria can still permit recurrent failures. We introduce \emph{average cost stability}, a criterion which is particularly useful for time-sensitive tasks, combining two observable quantities: the number of outstanding tasks and the cumulative number of irrecoverable deadline failures. Under bounded arrivals and uniformly bounded service windows, we show that the outstanding-task count is uniformly bounded independently of the assignment policy and prove that our average cost stability is equivalent to bounded expected cumulative failures. We further characterize \emph{degenerate backlog stability}, in which backlog remains bounded despite unbounded cumulative failures. We instantiate the framework in an adversarial pickup-and-delivery system where internal fleet agents spoof reported locations to attract assignments and leave requests unserviced. We develop deadline-aware assignment procedures and adversarial models with varying knowledge and coordination capabilities. Experiments using real mobility-on-demand request data and our proposed adversarial models demonstrate that backlog can remain bounded while cancellations persist, whereas our proposed average cost stability correctly identifies such behavior as unstable.

\end{abstract}

\begingroup
\renewcommand{\abstractname}{Note to Practitioners}
\begin{abstract}
Many automated dispatch and task-assignment systems are monitored primarily through queue length, utilization, or average service rates. These measures can be misleading when tasks have deadlines. A system may maintain a bounded queue while still allowing requests to expire repeatedly, creating the appearance of stable operation despite persistent customer-facing failures.

This paper provides a complementary reliability test based on quantities that operators can usually observe directly: the number of tasks still awaiting service and the total number that have missed their deadlines. The main practical implication is that operators should not treat bounded backlog as sufficient evidence of reliable operation. Repeated deadline failures should be tracked separately, particularly when assignment decisions depend on remotely reported agent states or when robots, vehicles, or communication systems may behave unpredictably.

The proposed test is intentionally strict. It is designed to identify whether deadline failures can continue indefinitely, not to determine whether a small failure rate is commercially acceptable. It should therefore be used alongside standard measures such as average waiting time, throughput, cancellation rate, and operating cost. The case study considers location manipulation in a transportation fleet, but the same monitoring approach could be adapted to warehouse orders, manufacturing jobs, inspection tasks, or service requests with expiration times. Practical deployment would require selecting meaningful deadlines and validating the test over operating horizons appropriate to the application.
\end{abstract}


\section{Introduction}

Automated task-assignment systems play a critical role in transportation, logistics, warehouse automation, manufacturing, robotic inspection, and service operations. In these systems, a centralized or distributed decision-making mechanism repeatedly allocates tasks to autonomous agents under stochastic demand and changing operating conditions. Examples include assigning transportation requests to autonomous vehicles, dispatching warehouse robots to service customer orders, routing automated guided vehicles to material-handling tasks, and allocating inspection or maintenance requests to mobile robots \cite{wollenstein2021routing,bai2022group,fernando2023graph,wilde2024statistically}. In many such applications, eventual task completion is not sufficient. Tasks must be initiated or completed within finite operational deadlines, after which they may be canceled, become infeasible, or lose their practical value.

A central question in these systems is how to define policy stability. Classical queueing-theoretic notions typically characterize stability through bounded workload, bounded queue length, or the ability of the available service capacity to keep pace with incoming demand \cite{spieser2014,zhang2018analysis,levin2022general}. These notions are well suited to systems in which unfinished tasks remain serviceable indefinitely and eventual completion is the primary objective. However, in deadline-constrained systems a bounded backlog does not necessarily imply reliable operation. A policy may maintain a finite but persistently large number of delayed tasks, or repeatedly allow tasks to expire while the number of currently outstanding tasks remains bounded. Indeed, when every task has a uniformly bounded lifetime, the number of outstanding tasks may be bounded by construction, since each task must eventually either be serviced or leave the system through expiration. Backlog boundedness can therefore become too weak to distinguish satisfactory operation from persistent deadline failure.

Conventional long-run average immediate cost criteria address a related but weaker reliability objective. For example, a stage cost based on the number of newly failed tasks at each time step measures the long-run average failure rate. Such a criterion is useful for comparing steady-state service quality across policies, but it can remain finite even when failures recur indefinitely at a bounded rate. Discounted reinforcement-learning objectives \cite{bertsekas2022lessons,bertsekas2023coursenew} have a different limitation: because outcomes receive progressively less weight farther into the future, persistent long-horizon degradation may have limited influence on the resulting objective. Average immediate cost and discounted criteria remain valuable for policy optimization and performance comparison, but they do not directly answer the stronger reliability question considered here: \emph{does a certain policy limit irrecoverable deadline failures to a finite total number, or can such failures recur indefinitely over the system's operating horizon?}

To address this question, we introduce an \emph{average cost notion of stability} for deadline-constrained automated task-assignment systems. The proposed formulation is based on two observable system-level quantities: the number of outstanding tasks and the cumulative number of tasks that have irrecoverably failed after exceeding their deadlines. These quantities are derived from observable \emph{operational outcomes}, through which the task-management system can determine whether tasks remain pending, are successfully completed, or have failed, without requiring reconstruction of the actions executed by individual agents or the amount of useful service work they perform. This distinction is important when agent behavior, hardware condition, communication quality, or environmental disturbances make \emph{execution-level behavior}, such as the actions performed and service progress achieved by individual agents, difficult to observe or model accurately.

The cumulative-failure term is intentional. Once a task fails irrecoverably, that failure remains part of the operational history of the system and continues to contribute to the proposed policy cost. Consequently, a policy under which failures persist at any positive long-run rate is unstable under the proposed criterion, even if its backlog and average number of failures per time step remain bounded. The proposed average cost stability therefore requires the expected total number of irrecoverable deadline failures over the infinite operating horizon to remain finite. It is not proposed as a replacement for throughput or service-level metrics, which remain useful for ranking policies and quantifying the severity of degradation. Rather, it provides a complementary and deliberately stronger reliability quantification for applications in which the indefinite recurrence of failures is itself unacceptable.

Using this cost, and under bounded task arrivals and uniformly bounded task deadlines, we show that the number of outstanding tasks is uniformly bounded independently of the assignment policy. At any time, the outstanding set can contain only tasks that arrived within a finite preceding interval. We then prove that the proposed stability criterion holds if and only if the expected cumulative number of irrecoverably failed tasks remains uniformly bounded over time. The criterion thus has a direct operational interpretation: failures may occur transiently, but a average cost stable policy cannot continue to generate deadline failures indefinitely in expectation.

This characterization also exposes a limitation of backlog-only stability. Under finite deadlines, expiration can keep the number of outstanding tasks bounded even while irrecoverable failures continue to accumulate. We refer to this regime as \emph{degenerate backlog stability}: a bounded-backlog condition is satisfied even though the expected cumulative number of irrecoverable failures is unbounded. When tasks do not expire, a related but distinct ambiguity arises: a system may maintain a large but bounded population of severely delayed tasks and still satisfy a bounded-backlog criterion despite unacceptable service quality. Finite deadlines convert such excessive delay into an observable task failure, and the cumulative-failure term prevents persistent deadline violations from being hidden by a bounded backlog. Deadlines therefore play a structural role in the proposed framework rather than serving only as an application-specific modeling detail.

We study adversarial pickup-and-delivery routing as a demanding instantiation of this framework. Routing and assignment mechanisms commonly assume that agents report their states truthfully and execute assigned routes cooperatively. In real cyber-physical systems, however, compromised or strategic agents may misreport their locations, attract assignments, increase service delays, and interfere with the effective use of cooperative fleet resources \cite{prorok2021beyond,zhou2021multi,sathaye2022experimental, dasgupta2024unveiling,liu2020secure,chu2024multi,yang2023location}. When reassignment is permitted, repeated manipulation of reported states may also induce assignment churn and redirect cooperative agents away from previously targeted requests. Such behavior makes the fleet's effective service capacity and agent-level productive effort difficult to characterize, while task-level outcomes such as outstanding and canceled requests remain directly observable. Adversarial routing therefore provides a useful stress test for an outcome-based stability formulation defined through observable task-level outcomes.

In the considered application, internal adversarial agents spoof their reported locations to attract pickup requests and then intentionally leave those requests unserviced. Internal adversarial agents are fleet agents that intentionally deviate from the prescribed task-assignment protocol by failing to execute assigned service actions, while remaining indistinguishable from cooperative agents to the centralized dispatcher because they participate in the same assignment process and are not known a priori to be adversarial. By attracting requests that would otherwise be assigned to cooperative agents, these adversaries induce inefficient allocations, service delays, request cancellations, and degraded fleet utilization. The attack model is motivated by known vulnerabilities of Global Navigation Satellite System (GNSS)-based localization \cite{petit2014potential,yang2021secure,abrar2024gps,zhao2025secure}, where spoofed signals can produce large and potentially stealthy localization errors. Although location spoofing provides the motivating source of disruption, the stability formulation does not depend on a particular failure mechanism. The same criterion can be applied when deadline failures result from hardware faults, communication delays, sensing errors, stochastic overload, environmental disturbances, or other forms of non-cooperative behavior.

To instantiate the framework, we develop routing and spoofing models that capture different levels of adversarial information and coordination capabilities. The adversarial models range from agents that observe only outstanding request locations to coordinated adversaries with knowledge of the cooperative fleet members and locations as well as the centralized assignment mechanism. We consider greedy assignment, instantaneous assignment without reassignment (IA), and instantaneous assignment with reassignment (IA-RA), covering both commitment-based and reassignment-based coordination structures. We also introduce feasibility-aware assignment procedures that account for request expiration by restricting assignments to agent-request pairs capable of satisfying the corresponding pickup deadlines.

Our empirical study uses real-world San Francisco mobility-on-demand request data \cite{piorkowski2009crawdad}, to which we introduce an adversarial subset of agents to the fleet that follow the described adversarial models. The experiments first examine fully cooperative fleets under different deadline lengths and fleet-size choices. For IA and IA-RA, existing queueing-theoretic fleet-sizing conditions \cite{garces2024approximate} are used to select nominal cooperative fleet sizes, while the greedy fleet size is calibrated empirically because a comparable analytical condition is not currently available. We first illustrate the ambiguity of backlog-only assessment in the absence of request expiration and then examine the empirical signature of degenerate backlog stability under finite request windows. Across multiple assignment policies, adversarial knowledge models, and fleet compositions, the results show that the backlog may remain bounded even while cancellations persist, whereas in contrast the proposed criterion identifies such regimes as violating average cost stability.

The main contributions of this paper are as follows:
\begin{itemize}
    \item \textbf{Average cost stability for deadline-constrained task assignment:} We introduce an observable stability notion requiring the expected cumulative number of irrecoverable task failures to remain uniformly bounded over time. The criterion is deliberately stronger than bounded-backlog stability or a finite long-run average failure rate.
    \item \textbf{Theoretical characterization and degenerate backlog stability:} Under bounded arrivals and uniformly bounded task deadlines, we establish a policy-independent bound on the number of outstanding tasks and prove that average cost stability is equivalent to uniform boundedness of the expected cumulative number of failed tasks. We also identify a \emph{degenerate backlog-stable regime} in which a policy satisfies a bounded-backlog criterion even though its expected cumulative number of irrecoverable task failures is unbounded.
    \item \textbf{Adversarial routing models and deadline-aware assignment:} We instantiate the proposed stability formulation in pickup-and-delivery routing with location-spoofing adversarial agents embedded within the fleet. We develop feasibility-aware versions of greedy, IA, and IA-RA assignment for finite request deadlines, together with adversarial models representing different levels of information and coordination capabilities.
    \item \textbf{Empirical evaluation using real-world mobility demand:} Using San Francisco mobility-on-demand data, we evaluate the framework across multiple assignment policies, adversarial knowledge models, deadline lengths, and fleet compositions. The experiments illustrate that bounded backlog does not imply average cost stability when deadline violations persist.
\end{itemize}

\subsection{Paper Organization}

The remainder of the paper is organized as follows. Section~\ref{section:related_work} reviews related work on task-assignment stability, deadline-constrained routing, and resilient multi-agent systems. Section~\ref{section:problem_formulation} introduces a general model of deadline-constrained automated task assignment and defines the observable system-level quantities used throughout the paper. Section~\ref{section:stability_definitions} presents the proposed strong average cost stability criterion. Section~\ref{sec:theoretical_results} establishes its main theoretical properties, including its characterization in terms of expected cumulative task failures and the phenomenon of degenerate backlog stability.

Section~\ref{sec:adversarial_routing} instantiates the general framework in adversarial pickup-and-delivery routing and introduces the considered location-spoofing, information, and assignment models. Section~\ref{sec:time_feasible_task_allocation} develops the deadline-aware assignment procedures used in this instantiation and analyzes their computational complexity. Section~\ref{sec:empirical_results} evaluates the framework using real-world mobility-demand data from San Francisco. Section~\ref{sec:generalization} discusses the broader applicability, practical implications, and limitations of the proposed stability notion. Finally, Section~\ref{sec:conclusion} summarizes the main findings and outlines directions for future work.

\section{Related Work}
\label{section:related_work}

This paper is related to three principal areas:  (i) stability and reliability in automated task-assignment systems,  (ii) deadline-constrained routing and task assignment, and  (iii) resilience to adversarial and non-cooperative behavior in cyber-physical and multi-agent systems.
The first two areas motivate the proposed average cost stability notion, while the third provides the demanding adversarial routing instantiation studied in this paper.

\subsection{Stability and Reliability in Automated Task-Assignment Systems}

Stability analysis is commonly used to characterize whether an automated service system possesses sufficient capacity to accommodate its incoming demand over an extended operating horizon. In queueing and dynamic routing models, stability is typically expressed through bounded workload, bounded queue length, or the ability of the available service capacity to prevent unbounded backlog growth. Within autonomous mobility systems, these ideas have been used to derive fleet-sizing conditions, characterize service capacity, and determine whether stochastic transportation demand can be sustained without unbounded accumulation of outstanding requests
\cite{spieser2014,zhang2018analysis}.

More recent work has developed dispatch and routing policies with explicit throughput or queue-stability guarantees. Examples include maximum-stability dispatch rules for autonomous mobility systems and maximum-pressure-type policies for stabilizing queue lengths in networked transportation systems \cite{levin2022general,xu2024fms,xu2024smoothing}. Related formulations have also been used to analyze online pickup-and-delivery and reassignment policies \cite{garces2024approximate}. These approaches generally characterize system performance using workload, service effort, or the number of requests awaiting
service.

Backlog-based formulations are appropriate when unfinished tasks remain serviceable indefinitely and eventual completion is the principal objective. They are less informative, however, when tasks have finite operational lifetimes. If every task eventually leaves the outstanding set through either service or expiration, the number of outstanding tasks may remain bounded even when deadline failures continue indefinitely. In this setting, queue boundedness alone does not distinguish reliable operation from a system that repeatedly loses tasks.

Average immediate cost and discounted formulations provide alternative measures of long-run policy performance \cite{bertsekas2022lessons,bertsekas2023coursenew}. A long-run average immediate cost based on newly failed tasks, for example, characterizes the steady-state task failure rate and is useful for comparing policies. Such a cost can nevertheless remain finite when failures recur indefinitely at a bounded rate. Discounted criteria similarly support policy optimization but assign progressively less weight to outcomes occurring farther in the future. The proposed average cost criterion addresses a different reliability question: whether the expected total number of irrecoverable failures remains bounded over the infinite operating horizon. It is intended to complement, rather than replace, throughput and service-level measures.

Many existing stability analyses also rely on assumptions about the service work performed by individual agents. Such assumptions are natural in cooperative systems, where agents are expected to execute their assigned routes or tasks. However, when agents fail, deviate from their instructions, or act adversarially, effective service capacity may be difficult to determine without explicitly modeling lost or wasted work \cite{francos2025Proofs}. This motivates an outcome-based formulation defined using observable task-level quantities rather than inferred agent-level execution.

\subsection{Deadline-Constrained Routing and Task Assignment}

Finite service deadlines are central to many transportation, logistics, and automation applications. Vehicle-routing problems with time windows have therefore been studied extensively in operations research, where task deadlines constrain route feasibility and determine whether customer requests can be serviced
\cite{solomon1987algorithms,braysy2005vehicle,lespay2021case,li2023multitask,tresca2024matheuristic,fernandez2025cumulative,jiang2025data}. These studies have produced exact, heuristic, and learning-based methods for constructing routes that satisfy temporal constraints while minimizing travel, delay, or operating cost.

Dynamic routing formulations extend these models to settings in which requests arrive online and assignments must be updated as the system evolves. Dynamic vehicle-routing and dial-a-ride models incorporate time windows while allowing routing and assignment decisions to respond to newly observed demand \cite{psaraftis1988dynamic,yang2017dynamic,ulmer2017dynamic,rios2021recent}. Similar considerations arise in industrial automation. Automated guided vehicle systems and just-in-time material-delivery applications must repeatedly allocate transportation tasks while respecting pickup, delivery, or production deadlines \cite{smolic2009time,nishida2022dynamic}.

This literature demonstrates the practical importance of task deadlines and provides methods for enforcing them within routing and scheduling algorithms. Its primary emphasis, however, is generally on feasibility, solution quality, and operational efficiency. Deadline violations are often represented through penalties, rejected requests, or service-level measures rather than through a long-run stability property. Conversely, many theoretical stability analyses assume that requests remain available until eventual service and therefore do not explicitly model expiration.

The present work connects these perspectives by treating task expiration as a structural component of the stability formulation. Under bounded arrivals and uniformly bounded deadlines, expiration imposes a policy-independent bound on
the number of outstanding tasks. This observation makes backlog boundedness insufficient as a reliability certificate and shifts attention to the accumulation of irrecoverable task failures. The proposed formulation therefore uses deadlines not only as assignment constraints, but also as the mechanism that converts excessive delay into an observable operational failure.

\subsection{Resilience to Adversarial and Non-Cooperative Behavior in Cyber-Physical and Multi-Agent Systems}


Resilience to adversarial, compromised, and otherwise non-cooperative agents has become increasingly important in cyber-physical and multi-agent systems. Prior surveys and theoretical studies have identified vulnerabilities arising from compromised agents, malicious sensor inputs, false state reports, and adversarial communication strategies \cite{etesami2019dynamic,prorok2021beyond,zhou2021multi,gil2023physicality}. These disruptions may invalidate the cooperative assumptions underlying conventional coordination, estimation, and control
algorithms.

In transportation and vehicular systems, false-data-injection attacks can manipulate traffic or state information and thereby induce congestion, inefficient routing, or unsafe decisions \cite{zhu2015game,eghtesad2026adversarial}. Localization systems present a particularly important attack surface. Spoofing or perturbing positioning signals can cause agents to report incorrect locations, disrupt navigation, and alter coordination decisions \cite{xu2023sok,li2024advgps,xu2025ghost}. A substantial portion of this literature focuses on detecting corrupted measurements, estimating the system state in the presence of compromised agents, or assigning trust values to agents and
sensors \cite{lei2025resilient,yemini2021characterizing}.

Adversarial behavior has also been studied in multi-robot task allocation. Recent resilient pickup-and-delivery frameworks, incorporate defensive mechanisms intended to reduce the effect of strategic or compromised
agents \cite{gong2025resilient}. This body of work primarily emphasizes attack detection, resilient assignment, and performance recovery. The long-run reliability consequences of adversarial interference, particularly when attackers induce the system to assign them tasks and subsequently leave them unserviced, have received less attention.

Adversarial pickup-and-delivery routing is used in this paper as a stress test for the proposed outcome-based stability formulation. Internal adversarial agents within the fleet may spoof their reported locations, attract assignments, and fail to perform the associated service. Under reassignment, they may also repeatedly trigger the reassignment of cooperative agents from one request to another, causing them to alter their routes.

In this setting, the amount of productive work performed by the fleet can be difficult to reconstruct from observable data because adversarial agents may appear to execute assignments while intentionally failing to complete them, whereas the numbers of outstanding and expired requests remain observable. This application therefore illustrates why task-level outcomes can provide a more robust basis for stability assessment when agent execution cannot be trusted.

\subsection{Positioning of This Work}

The reviewed literature provides several complementary perspectives on automated task-assignment systems. Queueing and routing research offers backlog- and workload-based notions of stability, but these notions primarily address whether service capacity can prevent unbounded task accumulation. Average immediate cost formulations characterize long-run service quality but can permit failures to continue indefinitely at a bounded rate. Deadline-constrained routing and scheduling research models expiration and temporal feasibility, but generally treats deadline violations as operational penalties rather than as the basis of a stability formulation. Research on adversarial multi-agent systems, meanwhile, has largely emphasized attack detection, resilient estimation, and mitigation.

This paper connects these areas by introducing an average cost notion of stability for deadline-constrained automated task assignment. The formulation is defined through observable system-level outcomes: outstanding tasks and cumulative irrecoverable task failures. Under bounded arrivals and uniformly bounded deadlines, we show that the number of outstanding tasks is uniformly bounded and that average cost stability is equivalent to uniform boundedness of the expected cumulative number of failed tasks. This distinguishes the proposed notion from bounded-backlog stability and from a finite average failure rate.

We further identify \emph{degenerate backlog stability}, in which a system satisfies a bounded-backlog condition while its expected cumulative number of irrecoverable task failures is unbounded. We then instantiate the framework in adversarial pickup-and-delivery routing with internal location-spoofing agents. This application demonstrates the value of an outcome-based stability criterion when agent-level execution and effective service effort cannot be reliably observed or modeled.

\section{Deadline-Constrained Task-Assignment Model}
\label{section:problem_formulation}

We consider a discrete-time automated task-assignment system operating over decision epochs \(t=0,1,2,\ldots\). A collection of agents performs tasks that arrive over time, while a decision-making mechanism repeatedly selects assignment, routing, scheduling, or control actions. The formulation is intentionally general and does not assume a particular physical environment, assignment algorithm, or source of performance degradation. The adversarial pickup-and-delivery problem considered later is introduced as a specific instantiation of this framework.

\subsection{Agents, Observed State, and Policies}

Let \( \mathcal{L}=\{1,\ldots,N\} \) denote the set of agents. At each decision epoch \(t\), the decision-making mechanism observes a system state \( \hat{x}_t\).
The state \(\hat{x}_t\) contains the quantities available to the decision maker that are relevant for selecting the current decision, evaluating system performance, and characterizing the subsequent evolution of the system. These quantities may include, for example, observed agent states, outstanding-task information, task attributes and deadlines, resource availability, and other application-specific variables.

The observed state need not coincide with an underlying physical state of the system. In particular, the quantities contained in \(\hat{x}_t\) may reflect noisy measurements, delayed or imperfect reports, execution uncertainty, hardware failures, or adversarially manipulated observations. The formulation only requires that \(\hat{x}_t\) be the state available to the decision-making mechanism at epoch \(t\).

A policy
\(
\pi=\{\mu_0,\mu_1,\ldots\}
\)
is a sequence of decision rules. At time \(t\), the policy selects an action,
\begin{equation}
    u_t
    =
    \mu_t(\hat{x}_t)
    \in
    \mathcal{U}_t(\hat{x}_t)
    \label{eq:general_policy}
\end{equation}
where \(\mathcal{U}_t(\hat{x}_t)\) denotes the set of admissible decisions when the observed state is \(\hat{x}_t\).

\subsection{State Evolution and Random Perturbations}

The evolution of the observed state is represented abstractly by,
\begin{equation}
    \hat{x}_{t+1}
    =
    \hat{f}_t(\hat{x}_t,u_t,\xi_t)
    \label{eq:general_state_dynamics}
\end{equation}
where $\xi_t$
is a random perturbation representing the exogenous stochasticity affecting the system during the transition from epoch \(t\) to epoch \(t+1\).

Depending on the application, \(\xi_t\) may contain stochastic task arrivals and their attributes, random travel or service times, execution disturbances, measurement errors, component failures, or other random quantities affecting the system evolution. Thus, conditional on the current observed state \(\hat{x}_t\) and selected action \(u_t\), the randomness in the next state is represented through \(\xi_t\).

The representation in \eqref{eq:general_state_dynamics} is intentionally general. In the routing application developed later, for example, the observed state includes the reported agent states and the current request information, while the perturbation contains the stochastic quantities governing new request arrivals and other exogenous uncertainty.

\subsection{Task Arrivals and Service Deadlines}

Let \(\mathcal{R}_t^{\mathrm{new}}\) denote the set of tasks entering the system at time \(t\), and let
 \(\eta_t=\left|\mathcal{R}_t^{\mathrm{new}}\right|\) 
denote the number of new arrivals. The arrival process and the attributes of the arriving tasks may be stochastic and can therefore be represented as components of the random perturbation \(\xi_t\).

Tasks may differ in their service requirements, locations, priorities, or other application-specific attributes. Each task \(r\) is represented abstractly by,
\begin{equation}
    r=\left\langle \theta_r,t_r,w_r\right\rangle
    \label{eq:general_task_tuple}
\end{equation}
where \(\theta_r\) contains the application-specific task attributes, \(t_r\) is the task's arrival time, and \(w_r\in\mathbb{N}^{+}\) is its allowable service window. The corresponding deadline is,
\begin{equation}
    d_r=t_r+w_r
    \label{eq:task_deadline}
\end{equation}
The attributes \(\theta_r\) specify the service milestone that must be satisfied by the deadline. Depending on the application, this milestone may correspond to initiating service, completing service, reaching a pickup location, delivering an item, or completing an inspection.

We adopt the following timing convention. Task \(r\) arrives at the beginning of epoch \(t_r\) and may satisfy its required service milestone during epochs \(t_r,t_r+1,\ldots,d_r\). If the milestone is satisfied no later than the end of epoch \(d_r\), the task is classified as successfully serviced. Otherwise, it becomes irrecoverably failed at the beginning of epoch \(d_r+1\). Once a task is classified as serviced or failed, that status becomes unmutable.

Let \(\mathcal{R}_{0:t}
=
\bigcup_{k=0}^{t}\mathcal{R}_k^{\mathrm{new}}
\)
denote the set of all tasks that have entered the system by time \(t\). This set can be represented as the union of three disjoint sets:
\begin{equation}
    \mathcal{R}_{0:t}
    =
    \mathcal{R}_t^{\mathrm{out}}
    \,\cup\,
    \mathcal{R}_t^{\mathrm{ser}}
    \,\cup\,
    \mathcal{R}_t^{\mathrm{fail}}
    \label{eq:task_status_partition}
\end{equation}
where
\begin{itemize}
    \item \(\mathcal{R}_t^{\mathrm{out}}\) is the set of tasks that have
    arrived but have not yet been successfully serviced or irrecoverably
    failed;

    \item \(\mathcal{R}_t^{\mathrm{ser}}\) is the cumulative set of tasks
    that have successfully satisfied their required service milestones; and

    \item \(\mathcal{R}_t^{\mathrm{fail}}\) is the cumulative set of tasks
    that have irrecoverably failed after exceeding their deadlines.
\end{itemize}

Because the task-status information is available to the task-management
mechanism, these sets are components of, or can be determined from, the
observed state \(\hat{x}_t\).

The proposed stability formulation therefore requires only that the observed state contain sufficient task-status information to determine
\(\mathcal{R}_t^{\mathrm{out}}\) and
\(\mathcal{R}_t^{\mathrm{fail}}\). It does not require the system operator to determine why an individual task failed or to reconstruct the physical actions actually performed by each agent.

In the adversarial pickup-and-delivery instantiation developed later, \(\theta_r\) contains the pickup and drop-off locations of request \(r\), and the required deadline milestone is the request pickup event. Accordingly, outstanding tasks correspond to requests awaiting pickup, whereas irrecoverably failed tasks correspond to requests canceled after exceeding their pickup deadlines.

\section{Average Cost Stability}
\label{section:stability_definitions}

We now introduce a new stability notion that distinguishes policies under which irrecoverable task failures remain finite from policies under which failures continue throughout the operating horizon. The proposed criterion is deliberately stronger than bounded-backlog stability and conventional long-run average failure-rate measures.

\subsection{Average Policy Cost}

We define the stage cost at time \(t\) as,
\begin{equation}
    g_t(\hat{x}_t, u_t, \xi_t)
    =
    |\mathcal{R}_t^{\mathrm{out}}|+ |\mathcal{R}_t^{\mathrm{fail}}|
    \label{eq:finite_failure_stage_cost}
\end{equation}
where \(|\mathcal{R}_t^{\mathrm{out}}|\) is the number of outstanding tasks and \(|\mathcal{R}_t^{\mathrm{fail}}|\) is the cumulative number of irrecoverably failed tasks. The number of outstanding and failed tasks depends on the task arrival process contained inside $\xi_t$ and the actions executed by the fleet contained in $u_t$.

Both terms are observable at the task-management level. The first represents unfinished work currently remaining in the system. The second retains every previous deadline failure as part of the system's operational history. A failure therefore does not disappear from the criterion merely because the associated task has left the outstanding set.

One could define a finite-failure condition directly by requiring the cumulative failure count to remain bounded. We instead introduce the policy cost in \eqref{eq:finite_failure_policy_cost} for two reasons. First, retaining the outstanding-task term allows the criterion to detect unbounded unfinished work in settings where task windows are not uniformly bounded. Second, the cost provides a scalar system-level functional that can be used to evaluate policies and may support policy-improvement methods in future work. The theoretical results in Section~\ref{sec:theoretical_results} show that, under appropriate regularity conditions, this policy cost admits an exact interpretation in terms of cumulative task failures.

For an observed state $\hat{x}_t$ at an initial time $t$, we define the average policy cost as,
\begin{equation}
\begin{split}
    J_{\pi}^{\mathrm{avg}}(\hat{x}_t)
    =
    \limsup_{T\rightarrow\infty}
    \frac{1}{T}
    \mathbb{E}
    \left[
        \sum_{k=t}^{T}
        g_k(\hat{x}_k, \mu_k(\hat{x}_k),\xi_k)
    \right]
\end{split}
\label{eq:finite_failure_policy_cost}
\end{equation}
The expectation is taken with respect to the stochastic evolution induced by
the perturbation process and policy \(\pi\), conditional on the observed
initial state \(\hat{x}_t\).

Although \eqref{eq:finite_failure_policy_cost} contains a time average, it is not a conventional average failure-rate objective. In \eqref{eq:finite_failure_policy_cost}, the quantity \(|\mathcal{R}_t^{\mathrm{fail}}|\) is cumulative, so every irrecoverable failure contributes to every subsequent stage cost. This persistent accounting gives the criterion its strong finite-failure interpretation.

\begin{definition}[Average Cost Stability]
\label{def:strong_finite_failure_stability}
A policy $\pi$ is stable if $J_\pi^{\mathrm{avg}}(\hat{x}_t)<\infty$, for all states $\hat{x}_t$ for all $t$.
\end{definition}

The use of cumulative failures is intentional. If the expected cumulative failure count grows without bound, past failures increasingly affect the average policy cost, even when their accumulation is sublinear and the long-run average failure rate is zero. Average cost stability therefore excludes policies under which the expected total number of irrecoverable deadline violations diverges.

The definition does not require failure-free operation. An average cost stable policy may incur transient deadline violations, and the number and timing of those violations may vary across realizations. As shown in Section~\ref{sec:theoretical_results}, under the stated regularity conditions, the criterion is equivalent to requiring a finite expected total number of irrecoverable failures. This further implies that the total number of failures is finite almost surely.

The criterion does not determine whether a particular finite number of failures is commercially or operationally acceptable. It should therefore complement rather than replace measures such as average delay, throughput, service level, and operating cost.

\begin{remark}[Positive weighting]
\label{rem:positive_finite_failure_weights}
The unit weights in \eqref{eq:finite_failure_stage_cost} are adopted for
notational simplicity. More generally, one may define
\begin{equation}
    g_t(\hat{x}_t,u_t,\xi_t)
    =
    \alpha |\mathcal{R}_t^{\mathrm{out}}|+\beta |\mathcal{R}_t^{\mathrm{fail}}|,
    \qquad
    \alpha,\beta>0
    \label{eq:weighted_finite_failure_stage_cost}
\end{equation}
Any fixed positive weights yield the same finite-versus-infinite stability
classification, although they change the numerical value of the policy cost
and the relative emphasis placed on unfinished and failed tasks.
\end{remark}

The definition above does not by itself imply that the outstanding-task count is bounded. In Section~\ref{sec:theoretical_results}, we introduce uniform bounds on task arrivals and service-window lengths and show that these conditions bound the number of outstanding tasks under every policy. Under the same conditions, average policy stability is equivalent to uniform boundedness of the expected cumulative failure count.

\section{Theoretical Characterization}
\label{sec:theoretical_results}

The preceding sections introduced a general deadline-constrained task-assignment model and defined average cost stability without imposing uniform restrictions on the task-arrival process or service-window
lengths. We now introduce regularity conditions under which the proposed criterion admits an exact operational characterization.

We first show that bounded arrivals and uniformly bounded service windows impose a policy-independent bound on the number of outstanding tasks. We then prove that, under these conditions, average cost stability is equivalent to a finite expected total number of irrecoverable task failures. Finally, we show that the same conditions make conventional backlog boundedness non-discriminating: every policy has bounded backlog, including policies under which deadline failures continue indefinitely.

\subsection{Regularity Conditions and Outstanding-Task Bound}

We begin with two assumptions governing task arrivals and service windows.

\begin{assumption}[Uniformly bounded arrivals]
\label{assump:bounded_arrivals}
There exists a finite constant $\bar{\eta}>0$ such that $\eta_t
    \leq
    \bar{\eta}$
for every $t$.
\end{assumption}

\begin{assumption}[Uniformly bounded service windows]
\label{assump:bounded_task_windows}
There exists a finite constant
$\bar{w}\in\mathbb{N}^{+}$ such that $w_r
    \leq
    \bar{w}$
for every request $r$.
\end{assumption}

Assumption~\ref{assump:bounded_arrivals} limits the number of tasks that can enter the system during a single decision epoch. Assumption~\ref{assump:bounded_task_windows} requires a common finite upper bound on the amount of time that any task may remain serviceable. These conditions do not require arrivals to be independent, identically distributed, or stationary.

We will use the following immediate consequence of Assumptions~\ref{assump:bounded_arrivals}
and~\ref{assump:bounded_task_windows}.

\begin{remark}[Uniform boundedness of outstanding requests]
\label{rem:bounded_outstanding_requests}
Under Assumptions~\ref{assump:bounded_arrivals}
and~\ref{assump:bounded_task_windows}, the number of outstanding requests is uniformly bounded over time. In particular, since each request can remain outstanding for at most $\bar{w}$ time steps and at most $\bar{\eta}$ new requests arrive at each time step, we have that:
\[
    |\mathcal{R}_t^{\mathrm{out}}| \le \bar{\eta}\bar{w}
\]
Thus, under bounded arrivals and uniformly finite request time windows, request backlog cannot grow without bound; only the cumulative number of canceled requests can cause unbounded growth in the proposed stage cost.
\end{remark}

\subsection{Characterization of Average Cost Stability}

Remark~\ref{rem:bounded_outstanding_requests} shows that finite request time windows prevent indefinite accumulation of outstanding requests. As a result, the outstanding-request term in the proposed stage cost is uniformly bounded independently of the policy. The next theorem uses this fact to show that stability under the proposed average cost criterion is equivalent to controlling the cumulative number of canceled requests.

\begin{theorem}[Characterization of Average Cost Stability]
\label{thm:main_stability_characterization}
Suppose Assumptions~\ref{assump:bounded_arrivals} and~\ref{assump:bounded_task_windows} hold. Given a policy $\pi$, $J_{\pi}^{\mathrm{avg}}(\hat{x}_t) < \infty$ for all $\hat{x}_t$ and all $t$ if and only if 
\begin{equation}
    \sup_{k\geq 0} \mathbb{E}[|\mathcal{R}_k^{\mathrm{fail}}|]
    <
    \infty
    \label{eq:theorem_finite_total_failures}
\end{equation}
\end{theorem}

\begin{proof}
Let \(O_k := |\mathcal{R}_k^{\mathrm{out}}|\), \(F_k := |\mathcal{R}_k^{\mathrm{fail}}|\), and \(B_{\mathrm{out}} := \bar{\eta}\bar{w} \). By Remark~\ref{rem:bounded_outstanding_requests}, we know that \(O_k
    \leq
    B_{\mathrm{out}}
    \qquad
    \text{for every }k
\) for any policy.

Moreover, since $F_k$ denotes the cumulative number of canceled requests by time $k$, the sequence $\{F_k\}_{k\ge 0}$ is positive and nondecreasing. Hence $\{\mathbb{E}[F_k]\}_{k\ge 0}$ is also positive and nondecreasing.

We now prove both directions.

\medskip
\noindent\textbf{($\Rightarrow$) Finite average policy cost implies bounded expected cumulative failed requests.}

Assume that
\[
    J_\pi^{\mathrm{avg}}(\hat{x}_t)<\infty 
\]
for all observed states $\hat{x}_t$. We show that this implies,
\[
    \sup_{k\ge 0}\mathbb{E}[F_k]<\infty 
\]

Suppose, for contradiction, that,
\[
    \sup_{k\ge 0}\mathbb{E}[F_k]=\infty 
\]
Since $\mathbb{E}[F_k]$ is positive and nondecreasing in $k$, for any constant $M>0$, there exists some time $K\ge t$ such that,
\[
    \mathbb{E}[F_k]\ge M
    \qquad \text{for all } k\ge K 
\]
Because the stage cost satisfies,
\[
    g\left(\hat{x}_k,\mu_k(\hat{x}_k),\xi_k\right)
    =
    O_k + F_k
    \ge F_k
\]
we obtain, for all $T\ge K$,
\[
\begin{aligned}
    \frac{1}{T}
    \mathbb{E}\left[
        \sum_{k=K}^{T}
        g\left(\hat{x}_k,\mu_k(\hat{x}_k),\xi_k\right)
    \right]
    &\ge
    \frac{1}{T}
    \sum_{k=K}^{T}
    \mathbb{E}[F_k]  \\
    &\ge
    \frac{1}{T}
    \sum_{k=K}^{T} M \\
    &=
    M\frac{T-K+1}{T}
\end{aligned}
\]
Taking the limit superior as $T\to\infty$ gives,
\[
    J_\pi^{\mathrm{avg}}(\hat{x}_K)
    \ge M 
\]
Since $g\left(\hat{x}_k,\mu_k(\hat{x}_k), \xi_k\right)$ is non-negative, we have that:
\begin{align*}
    & \frac{1}{T}
    \mathbb{E}\left[
        \sum_{k=t}^{T}
        g\left(\hat{x}_k,\mu_k(\hat{x}_k),\xi_k\right)
    \right] \\
    & \qquad \ge \frac{1}{T}
    \mathbb{E}\left[
        \sum_{k=K}^{T}
        g\left(\hat{x}_k,\mu_k(\hat{x}_k),\xi_k\right)
    \right]
\end{align*}
for all $t < K$. This implies that,
\[
    J_\pi^{\mathrm{avg}}(\hat{x}_t) \geq J_\pi^{\mathrm{avg}}(\hat{x}_K) \ge M
\]
for $t < K$. 
Since this relationship holds for any $M>0$, for a sufficiently large $M$, we have that,
\[
    J_\pi^{\mathrm{avg}}(\hat{x}_t)=\infty
\]
which contradicts the assumed finite average cost. Therefore,
\[
    \sup_{k\ge 0}\mathbb{E}[F_k]<\infty 
\]

\medskip
\noindent\textbf{($\Leftarrow$) Bounded expected cumulative failed requests imply finite average policy cost.}

Now assume that,
\[
    \sup_{k\ge 0}\mathbb{E}[F_k]<\infty 
\]
Then there exists a finite constant $B_{\mathrm{fail}}<\infty$ such that,
\[
    \mathbb{E}[F_k]\le B_{\mathrm{fail}}
    \qquad \text{for all } k 
\]
Using the uniform bound on outstanding requests from
Remark~\ref{rem:bounded_outstanding_requests}, we have,
\[
    \mathbb{E}[O_k]\le \bar{\eta}\bar{w}
    \qquad \text{for all } k 
\]
Therefore, for every $k$,
\[
\begin{aligned}
    \mathbb{E}\left[
        g\left(\hat{x}_k,\mu_k(\hat{x}_k),\xi_k\right)
    \right]
    &=
    \mathbb{E}[O_k+F_k] \\
    &=
    \mathbb{E}[O_k]+\mathbb{E}[F_k] \\
    &\le
    \bar{\eta}\bar{w}+B_{\mathrm{fail}} 
\end{aligned}
\]
Hence,
\[
\begin{aligned}
    J_\pi^{\mathrm{avg}}(\hat{x}_t)
    &=
    \limsup_{T\to\infty}
    \frac{1}{T}
    \mathbb{E}\left[
        \sum_{k=t}^{T}
        g\left(\hat{x}_k,\mu_k(\hat{x}_k),\xi_k\right)
    \right] \\
    &\le
    \limsup_{T\to\infty}
    \frac{1}{T}
    \sum_{k=t}^{T}
    \left(\bar{\eta}\bar{w}+B_{\mathrm{fail}}\right) \\
    &=
    \limsup_{T\to\infty}
    \frac{T-t+1}{T}
    \left(\bar{\eta}\bar{w}+B_{\mathrm{fail}}\right) \\
    &=
    \bar{\eta}\bar{w}+B_{\mathrm{fail}}
    <
    \infty 
\end{aligned}
\]
Thus,
\[
    J_\pi^{\mathrm{avg}}(\hat{x}_t)<\infty 
\]

Combining the two directions proves that
\[
    J_\pi^{\mathrm{avg}}(\hat{x}_t)<\infty
    \quad \Longleftrightarrow \quad
    \sup_{t\ge 0}\mathbb{E}\!\left[
        F_k
    \right]<\infty 
\]
This completes the proof.
\end{proof}

Theorem~\ref{thm:main_stability_characterization} gives the proposed criterion a direct operational meaning. Under bounded arrivals and uniformly bounded service windows, the outstanding-task component is automatically controlled. A policy is therefore average cost stable precisely when it incurs a finite expected total number of irrecoverable deadline failures.

Importantly, the result does not require failures to accumulate at a positive rate. Any unbounded growth in the expected cumulative failure count violates average cost stability, even if that growth is sublinear.

\begin{remark}[Positive weighting]
Theorem~\ref{thm:main_stability_characterization} also holds for the weighted stage cost in \eqref{eq:weighted_finite_failure_stage_cost} whenever $\alpha,\beta>0$. The positive backlog weight preserves the policy-independent upper bound on the outstanding-task contribution, while the positive cumulative-failure weight preserves the equivalence between finite policy cost and finite expected total failures.
\end{remark}

\subsection{Degenerate Backlog Stability}
Remark~\ref{rem:bounded_outstanding_requests} also exposes a limitation of conventional backlog-based stability in deadline-constrained systems. Under Assumptions~\ref{assump:bounded_arrivals} and~\ref{assump:bounded_task_windows}, every policy satisfies,
\[
    \sup_{t\geq 0}
    \mathbb{E}[|\mathcal{R}_t^{\mathrm{out}}|]
    \leq
    \bar{\eta}\bar{w}
    <
    \infty
\]
This includes policies that repeatedly allow tasks to expire. Backlog boundedness is therefore guaranteed by the expiration mechanism rather than by effective service.

\begin{definition}[Degenerate backlog stability]
\label{def:degenerate_backlog_stability}
A policy exhibits \emph{degenerate backlog stability} if it satisfies a
bounded-backlog condition,
\begin{equation}
    \sup_{t\geq 0}
    \mathbb{E}[|\mathcal{R}_t^{\mathrm{out}}|]
    <
    \infty
    \label{eq:degenerate_bounded_backlog}
\end{equation}
while its expected cumulative number of irrecoverable failures is unbounded:
\begin{equation}
    \sup_{t\geq 0}
    \mathbb{E}[|\mathcal{R}_t^{\mathrm{fail}}|]
    =
    \infty
    \label{eq:degenerate_unbounded_failures}
\end{equation}
\end{definition}

Under uniformly bounded service windows, any policy satisfying \eqref{eq:degenerate_unbounded_failures} exhibits degenerate backlog stability. Such a policy would be classified as stable by a backlog-only criterion despite repeatedly failing to satisfy task deadlines.

A simple example illustrates this distinction. Suppose one task enters the system at every epoch, every task has the same finite service window, and the policy never services any task. The number of outstanding tasks remains bounded by the number of arrivals that can occur within one service window, but one task eventually fails for every task that arrives. Hence,
\[
    \sup_{t\geq 0} |\mathcal{R}_t^{\mathrm{out}}|
    <
    \infty,
    \qquad
    |\mathcal{R}_t^{\mathrm{fail}}|
    \rightarrow
    \infty
\]
The policy is backlog stable but not average cost stable.

Finite deadlines play two related roles in the proposed framework. First, they prevent tasks from remaining indefinitely in the outstanding set, yielding the uniform bound in Remark~\ref{rem:bounded_outstanding_requests}. Second, they convert excessive delay into an observable and irreversible failure. The cumulative failure term then ensures that persistent deadline violations cannot be hidden by bounded backlog.

When tasks do not expire, excessive delays may instead appear as a large but bounded number of outstanding tasks. Such a system may also satisfy a backlog-only criterion despite providing unacceptable service. The adversarial routing experiments in Section~\ref{sec:empirical_results} illustrate both manifestations of this mismatch: without expiration, a large backlog may plateau, while with finite request windows, the backlog remains bounded even as cancellations continue to accumulate.

The theoretical results therefore establish the distinction underlying the remainder of the paper:
\[
    \text{bounded backlog}
    \;\not\Rightarrow\;
    \text{average cost stability}.
\]
The adversarial pickup-and-delivery instantiation developed next examines how this distinction arises when internal agents manipulate the assignment process and intentionally leave requests unserviced.

\section{Adversarial Pickup-and-Delivery Instantiation}
\label{sec:adversarial_routing}

The theoretical characterization in Section~\ref{sec:theoretical_results} applies to any deadline-constrained task-assignment system in which outstanding and irrecoverably failed tasks are observable. We now instantiate that framework in an adversarial pickup-and-delivery system.

This application provides a demanding setting in which to evaluate the proposed stability notion. A centralized scheduler assigns transportation requests using an observed state that includes agent-reported locations and request information. Internal adversarial agents may manipulate their reported locations so that they appear to be more favorable assignment candidates than nearby cooperative agents. After attracting assignments, these agents intentionally fail to execute the corresponding pickup service. Consequently, the physical service effort of individual agents may not be directly recoverable from the state observed by the scheduler. Outstanding and canceled requests, however, remain directly observable at the task-management level. The routing application therefore illustrates the value of assessing stability through observable task outcomes rather than through reconstructed physical agent behavior.

\subsection{Routing Environment and Fleet}

The transportation environment is represented by a directed graph \(\mathcal{G} = (\mathcal{V},\mathcal{E})\), where \(\mathcal{V}\) is the set of locations and \(\mathcal{E}\subseteq\mathcal{V}\times\mathcal{V}\) is the set of directed travel connections. For each node \(i\in\mathcal{V}\), let \(\mathcal{N}_i = \{j\in\mathcal{V}\mid(i,j)\in\mathcal{E}\}\) denote its set of outgoing neighbors. Traversing any edge requires one decision epoch. The shortest-path travel time between nodes \(i,j\in\mathcal{V}\) is denoted by \(d_{\mathcal{G}}(i,j)\). 

The agent set \(\mathcal{L}\) introduced in Section~\ref{section:problem_formulation} is partitioned as, 
\begin{equation}
    \mathcal{L}
    =
    \mathcal{C}
    \cup
    \mathcal{A}
    \label{eq:fleet_partition}
\end{equation}
where \(\mathcal{C}\) and \(\mathcal{A}\) denote the cooperative and adversarial agents, respectively. The total fleet size is \(N = |\mathcal{C}|+|\mathcal{A}|\), and the adversarial fleet fraction is,
\begin{equation}
    p_{\mathrm{adv}}
    =
    \frac{|\mathcal{A}|}{N}
    \label{eq:adversarial_fleet_fraction}
\end{equation}
The partition is fixed during an operating run but is unknown to the centralized scheduler.

Each cooperative agent follows the assignments and routes prescribed by the scheduler. An adversarial agent remains a registered member of the fleet but may manipulate its reported location and decline to execute assigned pickup service. The centralized scheduler does not employ an attack-detection or trust-monitoring mechanism in the experiments considered here. The resulting setting is therefore referred to as an \emph{unmonitored adversarial assignment system}.

\subsection{Pickup-and-Delivery Requests}

Each transportation request \(r\) is represented by,
\begin{equation}
    r
    =
    \left\langle
        \rho_r,
        \delta_r,
        t_r,
        w_r
    \right\rangle
    \label{eq:routing_request_tuple}
\end{equation}
where \(\rho_r,\delta_r\in\mathcal{V}\) are the pickup and drop-off locations, \(t_r\) is the request-arrival time, and \(w_r\in\mathbb{N}^{+}\) is the maximum allowable waiting time before pickup. The corresponding pickup deadline is
\(
    d_r=t_r+w_r.
\)
Thus, relative to the general task representation
in~\eqref{eq:general_task_tuple}, the application-specific attributes are
\(
    \theta_r=(\rho_r,\delta_r),
\)
and the required deadline milestone is successful pickup.

Consistent with the timing convention introduced in Section~\ref{section:problem_formulation}, request \(r\) may be picked up during any epoch from \(t_r\) through \(d_r\). If an agent completes the pickup no later than the end of epoch \(d_r\), the request is classified as successfully serviced for purposes of the deadline-reliability criterion. If the pickup has not occurred by that time, the request is canceled and enters the failed set at the beginning of epoch \(d_r+1\).

Let \(\phi_{r,t} \in \{0,1\} \) 
indicate whether request \(r\) has been successfully picked up by time \(t\). Before pickup or expiration, the request belongs to the outstanding set \(\mathcal{R}_t^{\mathrm{out}}\). Upon pickup, it enters \(\mathcal{R}_t^{\mathrm{ser}}\) for purposes of the deadline criterion, even though the assigned cooperative agent may continue carrying the request toward its destination \(\delta_r\). If the pickup deadline expires first, the request enters the canceled-request set \(\mathcal{R}_t^{\mathrm{can}}\).

The routing-specific cancellation notation is related to the general
failure notation by $\mathcal{R}_t^{\mathrm{can}} = \mathcal{R}_t^{\mathrm{fail}}$.
Hence, an irrecoverable task failure in the general model corresponds exactly to a request cancellation in the pickup-and-delivery system.

Requests arrive stochastically over time. As in the general formulation, let \(\mathcal{R}_t^{\mathrm{new}}\) denote the set of requests arriving at epoch \(t\), with
\(
    \eta_t
    =
    \left|\mathcal{R}_t^{\mathrm{new}}\right|
\)
The random perturbation \(\xi_t\) introduced in \eqref{eq:general_state_dynamics} contains the exogenous stochastic quantities governing the routing system during the transition from epoch \(t\) to epoch \(t+1\). In particular, it includes the stochastic arrival of new requests and their associated attributes. The empirical distributions used to generate request arrivals, pickup locations, and drop-off locations are specified in Section~\ref{sec:empirical_results}. The theoretical results require only the bounded-arrival and bounded-window conditions in Assumptions~\ref{assump:bounded_arrivals} and~\ref{assump:bounded_task_windows}.

\subsection{Observed State and Centralized Assignment}

Let \( \boldsymbol{\nu}_t = [\nu_t^1,\ldots,\nu_t^N] \) denote the physical agent locations at time \(t\), and let \(\boldsymbol{\hat{\nu}}_t = [\hat{\nu}_t^1,\ldots,\hat{\nu}_t^N] \) denote the corresponding locations reported to the centralized scheduler.

For cooperative agents, the reported and physical locations coincide:
\begin{equation}
    \hat{\nu}_t^\ell
    =
    \nu_t^\ell,
    \qquad
    \ell\in\mathcal{C}
    \label{eq:truthful_cooperative_reports}
\end{equation}
For adversarial agents, the reported location may differ from the physical
location.

The observed state \(\hat{x}_t\) introduced in Section~\ref{section:problem_formulation} therefore contains the scheduler-visible quantities required for assignment and performance evaluation. In this routing application, these quantities include the reported agent locations \(\boldsymbol{\hat{\nu}}_t\), the current request sets and their attributes, current assignments, and the service status of agents. In particular, the task-management component of \(\hat{x}_t\) determines \(\mathcal{R}_t^{\mathrm{out}}\) and \(\mathcal{R}_t^{\mathrm{can}}\).

The observed state generally does not reveal whether a reported location is truthful. The scheduler therefore constructs its assignment decisions from \(\hat{x}_t\), rather than from the unobserved physical configuration \(\boldsymbol{\nu}_t\). This distinction provides the routing-specific interpretation of the general statement that the observed state need not coincide with the underlying physical state.

At each decision epoch, the scheduler selects an action according to,
\begin{equation}
    u_t
    =
    \mu_t(\hat{x}_t)
    \in
    \mathcal{U}_t(\hat{x}_t)
    \label{eq:routing_policy}
\end{equation}
where \(u_t\) contains the assignment and routing decisions prescribed at time \(t\). The admissible action set depends on the assignment structure, including whether previously assigned agents may be reassigned before pickup.

An agent is said to be \emph{assignable} at time \(t\) if it is eligible to receive a pickup assignment under the policy being used. At each decision time, the scheduler matches assignable agents to outstanding requests using the reported-location component of \(\hat{x}_t\). The corresponding agent-request assignment cost is,
\begin{equation}
    M_t(\ell,r)
    =
    d_{\mathcal{G}}
    \left(
        \hat{\nu}_t^\ell,
        \rho_r
    \right)
    \label{eq:reported_assignment_cost}
\end{equation}

Because requests have finite pickup deadlines, not every pair \((\ell,r)\) is necessarily feasible. Section~\ref{sec:time_feasible_task_allocation} introduces the deadline-feasibility tests used to exclude assignments whose predicted pickup travel time exceeds the remaining service window.

Together, the current observed state, the scheduler action, and the random perturbation determine the next observed state according to the routing
instance of
\begin{equation}
    \hat{x}_{t+1}
    =
    \hat{f}_t(\hat{x}_t,u_t,\xi_t)
    \label{eq:routing_state_dynamics}
\end{equation}
Here, \(\hat{f}_t\) accounts for cooperative agent movement, pickup and delivery progress, request-status updates, assignment changes, cancellations, and the arrival of new requests. Adversarially manipulated reports affect the reported-location component of the observed state but do not imply corresponding instantaneous changes in the physical locations of the
adversarial agents.

\subsection{Internal Location-Spoofing Adversaries}

The adversarial agents seek to disrupt service by attracting assignments that would otherwise be allocated to cooperative agents and subsequently leaving the corresponding requests unserviced. The attack therefore acts by corrupting a component of the scheduler's observed state rather than by directly modifying the assignment algorithm or the request-arrival process.

\begin{definition}[Internal location-spoofing model]
\label{def:non_monitored_spoofing_adversarial_models}
An adversarial agent \(\ell_a\in\mathcal{A}\) follows the internal location-spoofing model if it satisfies the following conditions:
\begin{enumerate}
    \item Before the scheduler selects \(u_t\) at each decision epoch, the agent may report an arbitrary location \(\hat{\nu}_t^{\ell_a}\in\mathcal{V}\).

    \item The agent retains its registered identity and does not create additional identities, impersonate another agent, modify request data, or alter reports submitted by other agents.

    \item The agent does not complete pickup service for requests assigned to it. Such requests remain in \(\mathcal{R}_t^{\mathrm{out}}\) until they are reassigned to and picked up by a cooperative agent or until their pickup deadlines expire.

    \item The centralized scheduler does not detect or correct the manipulated report and therefore treats \(\hat{\nu}_t^{\ell_a}\) as the agent's current location when constructing \(u_t=\mu_t(\hat{x}_t)\).
\end{enumerate}
\end{definition}

The arbitrary-location assumption models the effect of a successful localization-spoofing attack on the scheduler's observed state. It does not imply that an adversarial agent physically moves instantaneously between locations. Instead, the scheduler evaluates that agent as though it were located at the spoofed node.

The attack model isolates the effect of corrupted agent reports on centralized assignment. Adversarial agents do not create false requests, alter the road network, directly modify the random perturbation governing request arrivals, or physically prevent cooperative agents from moving. Their influence arises through their ability to manipulate the reported-location component of \(\hat{x}_t\), thereby changing the assignment action selected by the scheduler and occupying assignments without providing pickup service.

\subsection{Adversarial Knowledge Models}

We consider three adversarial knowledge models that differ in the scheduler-visible system quantities available to adversarial agents when they select their reported locations. These models characterize the information available to the adversaries and should be distinguished from the observed state \(\hat{x}_t\) available to the centralized scheduler.

\begin{enumerate}
     \item \textbf{Request-Only Model:} Adversarial agents observe the pickup locations of outstanding requests but do not know the identities or locations of the other adversarial or cooperative agents. Each adversary therefore selects its reported location using its local state and the current outstanding-request set.

    \item \textbf{Team-Information Model:} Adversarial agents observe the outstanding-request set and know the identities, current states, and selected targets of the other adversarial agents. This additional knowledge permits coordinated targeting and reduces redundant attempts by multiple adversaries to capture the same
    request.

    \item \textbf{Full-Information Model:} Adversarial agents observe the outstanding-request set, know the current reported states and identities of all cooperative and adversarial agents, and know the assignment policy and matching procedure used by the centralized scheduler. They can therefore predict the current assignment action that would be selected from \(\hat{x}_t\) and strategically target requests that would otherwise be assigned to cooperative agents.

\end{enumerate}

The full-information model corresponds to the \emph{omnipotent} adversarial model used in the experimental figures. Here, "omnipotent" refers to complete knowledge of the current assignment problem. It does not provide knowledge of future realizations of the random perturbations \(\xi_t\), and in particular does not reveal future request arrivals.

These models form a hierarchy of adversarial knowledge. The request-only model represents uncoordinated disruption based on current demand information. The team-information model additionally permits coordination among adversarial agents. The full-information model further permits strategic interception of assignments that would otherwise be made to cooperative agents. The knowledge-dependent procedures used to convert these observations into spoofed location reports are developed in Section~\ref{sec:time_feasible_task_allocation}.

\subsection{Assignment Structures}

We examine three assignment structures that instantiate different decision rules \(\mu_t\) in the general policy \(\pi=\{\mu_0,\mu_1,\ldots\}\). The policies differ in how they coordinate the fleet and in whether assignments may be updated before pickup.

\subsubsection{Instantaneous Assignment with Reassignment}

Under instantaneous assignment with reassignment (IA-RA), assignable agents and outstanding requests are repeatedly matched by solving a minimum-cost bipartite assignment problem using the costs in \eqref{eq:reported_assignment_cost}. We denote the resulting policy by \(\pi^{\mathrm{IA\text{-}RA}}\).

A cooperative agent that has not yet picked up a request remains assignable, even when it is currently traveling toward a previously assigned pickup. Consequently, the action \(u_t\) may revise pre-pickup assignments at subsequent decision epochs in response to changes in the observed state, including new requests and updated agent reports. Once a cooperative agent picks up a request, its assignment becomes fixed and the agent proceeds toward
the corresponding drop-off location.

IA-RA provides the greatest responsiveness to changes in the observed state, but its reassignment structure also creates an additional attack surface. Repeated changes in adversarial reports can modify the assignment action across successive epochs, displace cooperative agents from previous pickup targets, and induce assignment churn.

\subsubsection{Instantaneous Assignment without Reassignment}

Under instantaneous assignment without reassignment (IA), the scheduler uses the same minimum-cost matching structure but commits each assignment until the corresponding pickup is completed or the request expires. We denote this policy by \(\pi^{\mathrm{IA}}\).

A cooperative agent assigned to a request continues toward the corresponding pickup location and is not assignable to another request until that pickup is completed. An adversarial agent assigned to a request likewise remains associated with that request until expiration. Because the adversarial agent does not execute pickup service, the captured request remains outstanding unless the policy permits some other mechanism for releasing the assignment. This commitment structure limits reassignment-induced disruption but reduces the scheduler's ability to respond to newly arriving requests.

\subsubsection{Greedy Assignment}

Under greedy assignment, denoted by \(\pi^{\mathrm{G}}\), assignable agents are processed sequentially and each is matched to its nearest currently unmatched request according to the reported travel cost in \eqref{eq:reported_assignment_cost}. Assignments remain fixed until pickup or expiration.

Greedy assignment avoids solving a global fleet-level matching problem, but it does not jointly optimize assignments across the fleet and may therefore produce inefficient allocations under high demand. In the adversarial setting, its primary vulnerability is persistent local request blocking rather than reassignment churn.

The three policies therefore instantiate different mappings from the observed state \(\hat{x}_t\) to the scheduler action \(u_t\). IA-RA permits strategic disruption through repeated rematching, whereas IA and greedy assignment restrict this mechanism through assignment commitment but allow requests captured by adversarial agents to remain blocked until expiration.

\subsection{Connection to Average Cost Stability}

For the pickup-and-delivery instantiation, the outstanding-task count \(|\mathcal{R}_t^{\mathrm{out}}|\) is the number of requests awaiting pickup and the cumulative failure count \(|\mathcal{R}_t^{\mathrm{fail}}|\) is exactly the cumulative cancellation count \(|\mathcal{R}_t^{\mathrm{can}}|\). Consequently, the stage cost in \eqref{eq:finite_failure_stage_cost} becomes
\begin{equation}
    g_t
    (\hat{x}_t,u_t,\xi_t)
    =
    |\mathcal{R}_t^{\mathrm{out}}|+|\mathcal{R}_t^{\mathrm{can}}|
    \label{eq:routing_finite_failure_stage_cost}
\end{equation}

As in the general model, the explicit arguments \((\hat{x}_t,u_t,\xi_t)\) identify the observed state, scheduler decision, and random perturbation associated with epoch \(t\). For the particular average cost criterion used here, \(|\mathcal{R}_t^{\mathrm{out}}|\) and \(|\mathcal{R}_t^{\mathrm{can}}|\) are determined from the observable request-status component of \(\hat{x}_t\). The action \(u_t\) and perturbation \(\xi_t\) affect future stage costs through their effect on the subsequent observed-state trajectory.

Using the cancellation-failure correspondence \(|\mathcal{R}_t^{\mathrm{can}}|=|\mathcal{R}_t^{\mathrm{fail}}|\), Theorem~\ref{thm:main_stability_characterization} implies that, under bounded request arrivals and uniformly bounded pickup windows, a routing policy is average cost stable if and only if
\begin{equation}
    \sup_{t\geq 0}
    \mathbb{E}[|\mathcal{R}_t^{\mathrm{can}}|]
    <
    \infty
    \label{eq:routing_finite_cancellations}
\end{equation}
Thus, average cost stability in the routing system is equivalent to requiring a finite expected total number of request cancellations.

The empirical evaluation therefore does not treat bounded request backlog alone as evidence of reliable operation. Instead, it examines whether cancellations continue to accumulate under different assignment policies, adversarial knowledge models, and fleet compositions. This directly instantiates the distinction established in Section~\ref{sec:theoretical_results}: finite pickup windows can keep the number of outstanding requests bounded even when adversarial interference causes request cancellations to continue indefinitely.

The present section specifies the routing environment, the observed state and scheduler actions, the adversarial attack surface, and the policy classes used in that evaluation. The next section develops the deadline-feasible matching procedures and knowledge-dependent spoofing rules used to implement these policies under finite pickup windows.

\section{Deadline-Aware Assignment and Computational Complexity}
\label{sec:time_feasible_task_allocation}

The adversarial routing model in Section~\ref{sec:adversarial_routing} specifies which agents and requests may participate in each assignment decision. Finite pickup windows impose an additional requirement: a request should be assigned only if the scheduler believes that the selected agent can reach its pickup location before the deadline. A distance-only assignment rule may otherwise allocate requests to agents that cannot complete pickup in time, causing cancellations even in the absence of adversarial behavior.

This section incorporates deadline feasibility into the three assignment structures considered in the paper. We first define feasible agent-request pairs and the corresponding maximum-cardinality minimum-cost matching problem. We then describe how the adversarial information models use this feasibility structure to select spoofing targets. Finally, we analyze the computational complexity of the resulting assignment and adversarial-targeting procedures.

\subsection{Deadline-Feasible Agent-Request Pairs}

Let \(\mathcal{L}_t^{\pi}\subseteq\mathcal{L}\) denote the set of agents that are assignable at epoch \(t\) under policy \(\pi\). The composition of this set depends on whether the policy permits reassignment and is specified in Section~\ref{subsec:integration_assignment_structures}.

For each outstanding request \(r\in\mathcal{R}_t^{\mathrm{out}}\), define its remaining pickup slack as,
\begin{equation}
    s_{r,t}
    =
    d_r-t
    =
    (t_r+w_r)-t
    \label{eq:remaining_pickup_slack}
\end{equation}
Under the timing convention of Section~\ref{section:problem_formulation}, request \(r\) may be picked up during epoch \(d_r\). Thus, an assignable agent \(\ell\) is perceived to be capable of reaching request \(r\) before expiration if,
\begin{equation}
    d_{\mathcal{G}}
    \left(
        \hat{\nu}_t^\ell,\rho_r
    \right)
    \leq
    s_{r,t}
    \label{eq:deadline_feasibility_condition}
\end{equation}

Because the scheduler does not know the fleet partition \(\mathcal{L}=\mathcal{C}\,\dot{\cup}\,\mathcal{A}\), it applies \eqref{eq:deadline_feasibility_condition} to every assignable agent using that agent's reported location. In particular, an adversarial agent that reports the pickup node of a request appears to have zero travel time and is therefore perceived as feasible, even if its true location is elsewhere.

For policy \(\pi\), define the set of perceived deadline-feasible pairs as,
\begin{equation}
\begin{split}
    \mathcal{E}_t^{\pi}
    =
    \Big\{
        (\ell,r)
        \,\Big|\,
        &\ell\in\mathcal{L}_t^{\pi},\;
        r\in\mathcal{R}_t^{\mathrm{out}},\\
        &d_{\mathcal{G}}
        (\hat{\nu}_t^\ell,\rho_r)
        \leq s_{r,t}
    \Big\}
\end{split}
\label{eq:feasible_agent_request_edges}
\end{equation}
The corresponding set of requests that appear reachable by at least one assignable agent is,
\begin{equation}
    \mathcal{R}_{t,\pi}^{\mathrm{feas}}
    =
    \left\{
        r\in\mathcal{R}_t^{\mathrm{out}}
        \,\middle|\,
        \exists \ell\in\mathcal{L}_t^{\pi}
        :
        (\ell,r)\in\mathcal{E}_t^{\pi}
    \right\}
    \label{eq:feasible_request_set}
\end{equation}
Requests outside \(\mathcal{R}_{t,\pi}^{\mathrm{feas}}\) are excluded from the current assignment decision. They remain outstanding until they become reachable at a later epoch, are assigned under an updated reported state, or reach their deadlines and are canceled.

\subsection{Maximum-Cardinality Minimum-Cost Assignment}

For each feasible pair, the reported travel cost is,
\begin{equation}
    c_t(\ell,r)
    =
    d_{\mathcal{G}}
    \left(
        \hat{\nu}_t^\ell,\rho_r
    \right)
    \label{eq:deadline_aware_assignment_cost}
\end{equation}
Let \(\mathfrak{M}(\mathcal{E}_t^{\pi})\) denote the set of all one-to-one matchings whose edges belong to \(\mathcal{E}_t^{\pi}\). We first maximize the number of assigned requests and then minimize travel cost among matchings of maximum cardinality.
Define,
\begin{equation}
    q_t^{\star}
    =
    \max_{\mathcal{M}\in
    \mathfrak{M}(\mathcal{E}_t^{\pi})}
    |\mathcal{M}|
    \label{eq:maximum_assignment_cardinality}
\end{equation}
The deadline-aware matching is then
\begin{equation}
\begin{split}
    \mathcal{M}_t^{\pi}
    \in
    \arg\min_{\mathcal{M}}
    \quad&
    \sum_{(\ell,r)\in\mathcal{M}}
    c_t(\ell,r)\\
    \text{s.t.}\quad&
    \mathcal{M}\in
    \mathfrak{M}(\mathcal{E}_t^{\pi}),\\
    &
    |\mathcal{M}|=q_t^{\star}
\end{split}
\label{eq:max_cardinality_min_cost_assignment}
\end{equation}

Problem~\eqref{eq:max_cardinality_min_cost_assignment} can be implemented using an auction algorithm \cite{bertsekas1988auction,bertsekas2024new}, a rectangular linear-assignment solver \cite{crouse2016implementing}, or an equivalent minimum-cost matching method. Infeasible edges may be omitted from the bipartite graph or represented by a sufficiently large sentinel cost. The sentinel must exceed the cost of every admissible matching so that an infeasible pair is never preferred to leaving a request unmatched.

Algorithm~\ref{alg:deadline_aware_assignment} summarizes the common deadline-aware assignment procedure. For IA and IA-RA, the operator \textsc{MaxCardMinCostMatch} solves \eqref{eq:max_cardinality_min_cost_assignment}. For greedy assignment, agents are processed sequentially and assigned to their nearest unmatched feasible requests.

\begin{algorithm}[t]
\caption{Deadline-aware centralized assignment}
\label{alg:deadline_aware_assignment}
\begin{algorithmic}[1]
\Require Graph \(\mathcal{G}\), assignable agents
\(\mathcal{L}_t^{\pi}\), outstanding requests
\(\mathcal{R}_t^{\mathrm{out}}\), reported locations
\(\boldsymbol{\hat{\nu}}_t\), assignment structure \(\pi\)
\Ensure Deadline-feasible matching \(\mathcal{M}_t^{\pi}\)

\State \(\mathcal{E}_t^{\pi}\gets\emptyset\)
\For{each \(\ell\in\mathcal{L}_t^{\pi}\)}
    \For{each \(r\in\mathcal{R}_t^{\mathrm{out}}\)}
        \State \(c_t(\ell,r)\gets
        d_{\mathcal{G}}(\hat{\nu}_t^\ell,\rho_r)\)
        \If{\(c_t(\ell,r)\leq d_r-t\)}
            \State
            \(\mathcal{E}_t^{\pi}
            \gets
            \mathcal{E}_t^{\pi}\cup\{(\ell,r)\}\)
        \EndIf
    \EndFor
\EndFor

\If{\(\pi=\pi^{\mathrm{G}}\)}
    \State
    \(\mathcal{M}_t^{\pi}
    \gets
    \Call{GreedyFeasibleMatch}
    {\mathcal{L}_t^{\pi},
     \mathcal{R}_t^{\mathrm{out}},
     \mathcal{E}_t^{\pi},
     c_t}\)
\Else
    \State
    \(\mathcal{M}_t^{\pi}
    \gets
    \Call{MaxCardMinCostMatch}
    {\mathcal{L}_t^{\pi},
     \mathcal{R}_t^{\mathrm{out}},
     \mathcal{E}_t^{\pi},
     c_t}\)
\EndIf
\State \Return \(\mathcal{M}_t^{\pi}\)
\end{algorithmic}
\end{algorithm}

\subsection{Cooperative-Feasibility Prediction by Full-Information Adversaries}

The server evaluates feasibility using the reports of all agents. A full-information adversary performs an additional calculation: it predicts which requests would be assigned to cooperative agents if the adversarial agents were absent. Targeting these requests has a direct operational effect because it displaces assignments that would otherwise result in legitimate service.

Let \(\mathcal{C}_t^{\pi}\subseteq\mathcal{C}\) denote the cooperative agents that would be assignable under policy \(\pi\). Because cooperative reports are truthful, define the cooperative-feasible edge set as,
\begin{equation}
\begin{split}
    \mathcal{E}_{t,\mathrm{coop}}^{\pi}
    =
    \Big\{
        (\ell,r)
        \,\Big|\,
        &\ell\in\mathcal{C}_t^{\pi},\;
        r\in\mathcal{R}_t^{\mathrm{out}},\\
        &d_{\mathcal{G}}
        (\nu_t^\ell,\rho_r)
        \leq d_r-t
    \Big\}
\end{split}
\label{eq:cooperative_feasible_edges}
\end{equation}
The full-information adversary solves the corresponding cooperative-only assignment problem and obtains a predicted matching,
\begin{equation}
    \mathcal{M}_{t,\mathrm{coop}}^{\pi}
    \label{eq:predicted_cooperative_matching}
\end{equation}
Its primary target set is therefore,
\begin{equation}
    \mathcal{R}_{t,\mathrm{target}}^{\mathrm{full}}
    =
    \left\{
        r
        \,\middle|\,
        (\ell,r)\in
        \mathcal{M}_{t,\mathrm{coop}}^{\pi}
    \right\}
    \label{eq:full_information_target_set}
\end{equation}

This simulation uses only current information. The full-information adversary knows the current fleet state, request set, and assignment mechanism, but it does not know future request arrivals or future stochastic disturbances.

\subsection{Knowledge-Dependent Adversarial Targeting}

Let \(\mathcal{A}_t^{\pi}\subseteq\mathcal{A}\) denote the adversarial agents that are assignable at epoch \(t\) under policy \(\pi\). Before the centralized assignment is solved, each such adversary selects a target request and reports that request's pickup location:
\begin{equation}
    \hat{\nu}_t^a
    =
    \rho_{r_a^{\mathrm{target}}},
    \qquad
    a\in\mathcal{A}_t^{\pi}
    \label{eq:adversarial_spoofed_pickup_location}
\end{equation}
The target-selection procedure depends on the information model.

Under the \emph{full-information model}, adversaries first target requests in \(\mathcal{R}_{t,\mathrm{target}}^{\mathrm{full}}\), since these requests are predicted to be assigned to cooperative agents. Distinct adversaries are assigned to distinct targets whenever possible. If more assignable adversaries than predicted cooperative targets are available, the remaining adversaries target additional outstanding requests using the coordinated team-information rule.

Under the \emph{team-information model}, adversaries coordinate over the current outstanding-request set. They compute a one-to-one matching between assignable adversaries and outstanding requests using,
\begin{equation}
    c_t^{\mathrm{adv}}(a,r)
    =
    d_{\mathcal{G}}
    \left(
        \nu_t^a,\rho_r
    \right)
    \label{eq:adversarial_targeting_cost}
\end{equation}
thereby selecting distinct request targets while reducing redundant targeting.

Under the \emph{request-only model}, adversaries do not coordinate. Each adversary independently selects the outstanding request nearest to its true location:
\begin{equation}
    r_a^{\mathrm{target}}
    \in
    \arg\min_{r\in\mathcal{R}_t^{\mathrm{out}}}
    d_{\mathcal{G}}
    \left(
        \nu_t^a,\rho_r
    \right)
    \label{eq:request_only_target}
\end{equation}
Multiple adversaries may therefore select the same request.

Algorithm~\ref{alg:knowledge_dependent_targeting} summarizes these procedures.

\begin{algorithm}[t]
\caption{Knowledge-dependent adversarial targeting}
\label{alg:knowledge_dependent_targeting}
\begin{algorithmic}[1]
\Require Graph \(\mathcal{G}\), assignable adversaries
\(\mathcal{A}_t^{\pi}\), assignable cooperative agents
\(\mathcal{C}_t^{\pi}\), outstanding requests
\(\mathcal{R}_t^{\mathrm{out}}\), true cooperative locations,
assignment structure \(\pi\), information model \(\mathcal{K}\)
\Ensure Spoofed reports
\(\hat{\nu}_t^a\) for all \(a\in\mathcal{A}_t^{\pi}\)

\If{\(\mathcal{R}_t^{\mathrm{out}}=\emptyset\)}
    \State \Return
\EndIf

\If{\(\mathcal{K}=\textsc{FullInfo}\)}
    \State
    \(\mathcal{M}_{t,\mathrm{coop}}^{\pi}
    \gets
    \Call{DeadlineAware}
    {\mathcal{G},
     \mathcal{C}_t^{\pi},
     \mathcal{R}_t^{\mathrm{out}},
     \boldsymbol{\nu}_t^{\mathcal{C}},
     \pi}\)
    \State
    \(\mathcal{R}_{t,\mathrm{target}}
    \gets
    \{r:(\ell,r)\in
    \mathcal{M}_{t,\mathrm{coop}}^{\pi}\}\)
    \State Assign distinct adversaries to distinct requests in
    \(\mathcal{R}_{t,\mathrm{target}}\)
    \State Assign any surplus adversaries to distinct requests in
    \(\mathcal{R}_t^{\mathrm{out}}
    \setminus\mathcal{R}_{t,\mathrm{target}}\)
    using the team-information rule

\ElsIf{\(\mathcal{K}=\textsc{TeamInfo}\)}
    \State Compute a maximum-cardinality minimum-cost matching between
    \(\mathcal{A}_t^{\pi}\) and
    \(\mathcal{R}_t^{\mathrm{out}}\) using
    \(c_t^{\mathrm{adv}}(a,r)\)

\ElsIf{\(\mathcal{K}=\textsc{RequestOnly}\)}
    \For{each \(a\in\mathcal{A}_t^{\pi}\)}
        \State
        \(r_a^{\mathrm{target}}
        \gets
        \arg\min_{r\in\mathcal{R}_t^{\mathrm{out}}}
        d_{\mathcal{G}}(\nu_t^a,\rho_r)\)
    \EndFor
\EndIf

\For{each adversary \(a\) with target
\(r_a^{\mathrm{target}}\)}
    \State
    \(\hat{\nu}_t^a
    \gets
    \rho_{r_a^{\mathrm{target}}}\)
\EndFor
\State \Return spoofed adversarial reports
\end{algorithmic}
\end{algorithm}

The reported positions generated by Algorithm~\ref{alg:knowledge_dependent_targeting} are supplied to the same centralized assignment mechanism used by the rest of the fleet. Thus, the adversaries do not directly choose the final matching. They manipulate the reported cost and feasibility structure so that the scheduler is more likely to allocate their targeted requests to them.

\subsection{Integration with Assignment Policies}
\label{subsec:integration_assignment_structures}

The deadline-feasibility test and adversarial-targeting procedures are integrated differently under IA-RA, IA, and greedy assignment because the policies define different sets of assignable agents.

\subsubsection{IA-RA}
Under IA-RA, every agent that is not currently transporting a picked-up request is assignable. Cooperative agents traveling toward a pickup remain eligible for reassignment, and their previous pre-pickup assignments may be replaced at the next decision epoch. Because adversarial agents never complete pickup, they also remain assignable at every epoch.

The centralized scheduler therefore reconstructs the feasible graph \(\mathcal{E}_t^{\mathrm{IA\text{-}RA}}\) and resolves \eqref{eq:max_cardinality_min_cost_assignment} at every decision epoch. This repeated optimization permits the scheduler to respond to new requests, but it also allows adversarial agents to change their targets and alter the matching repeatedly. IA-RA is consequently exposed to both request blocking and reassignment churn.

\subsubsection{IA}
Under IA, only uncommitted agents are assignable. Once a cooperative agent is matched to a request, its pre-pickup assignment remains fixed until pickup. After pickup, the agent transports the request to its drop-off location and becomes assignable again only after completing the trip. An adversarial agent that captures a request remains associated with that request until its pickup window expires, after which the adversary becomes assignable again.

Deadline feasibility is therefore evaluated only when a new commitment is created. Existing commitments are not reconsidered at later epochs. Relative to IA-RA, this structure prevents repeated rematching of the same cooperative agent but permits an adversary to block a captured request for the remainder of its service window.

\subsubsection{Greedy Assignment}
Greedy assignment uses the same commitment and assignability rules as IA, but does not solve a global matching problem. Each assignable agent is processed sequentially and matched to its nearest currently unmatched request satisfying \eqref{eq:deadline_feasibility_condition}. Once assigned, the agent remains committed until pickup and trip completion in the cooperative case or until request expiration in the adversarial case.

The greedy policy therefore avoids reassignment churn and global matching overhead. Its lack of fleet-level coordination, however, can lead to inefficient assignments and makes its outcome sensitive to the order in which agents are processed.

\begin{remark}[Surplus adversaries]
\label{remark:large_proportions_of_advs}
Under the full-information model, the primary adversarial targets are requests predicted to be assigned to cooperative agents. If
\[
    |\mathcal{A}_t^{\pi}|
    >
    |\mathcal{R}_{t,\mathrm{target}}^{\mathrm{full}}|
\]
the remaining adversaries target additional outstanding requests. These requests may not have been assigned during the predicted cooperative matching, but blocking them can prevent subsequent service as cooperative agents become available. Under IA and greedy assignment, a surplus adversary remains committed to the captured request until expiration. Under IA-RA, it may select a new target at the next decision epoch.
\end{remark}

\subsection{Computational Complexity}
\label{subsec:computational_complexity}

We now characterize the computational cost of deadline screening, centralized assignment, and adversarial target selection. Let
\begin{equation}
    n_t
    =
    |\mathcal{L}_t^{\pi}|,
    \qquad
    m_t
    =
    |\mathcal{R}_t^{\mathrm{out}}|,
    \qquad
    a_t
    =
    |\mathcal{A}_t^{\pi}|
    \label{eq:complexity_dimensions}
\end{equation}
denote the numbers of assignable agents, outstanding requests, and assignable adversaries at epoch \(t\), respectively. Let \(\kappa_t\) denote the largest
finite integer cost in the assignment matrix.

We assume that shortest-path distances between graph nodes are precomputed. Under this assumption, each distance query has constant cost. Constructing the feasible edge set in \eqref{eq:feasible_agent_request_edges} therefore requires
\begin{equation}
    O(n_t m_t)
    \label{eq:feasibility_screening_complexity}
\end{equation}
time and at most \(O(n_t m_t)\) storage. If distances are computed online, the cost of the chosen shortest-path algorithm must be added to these bounds.

\paragraph{Greedy assignment}
Each assignable agent may inspect up to \(m_t\) requests to identify its nearest unmatched feasible request. Its per-epoch complexity, including deadline screening, is therefore
\begin{equation}
    O(n_t m_t).
    \label{eq:greedy_complexity}
\end{equation}

\paragraph{IA and IA-RA}
For the auction implementation used in this work, a rectangular assignment between \(n_t\) agents and \(m_t\) requests has per-epoch complexity \cite{bertsekas2024new}
\begin{equation}
    O\!\left(
        n_t m_t
        \log(n_t\kappa_t)
    \right)
    \label{eq:auction_complexity}
\end{equation}
which dominates the \(O(n_t m_t)\) feasibility-screening cost. IA and IA-RA therefore have the same worst-case per-epoch asymptotic complexity. Their practical costs differ because IA optimizes only over newly assignable agents, whereas IA-RA repeatedly includes all agents whose requests have not yet been picked up.

\paragraph{Adversarial preprocessing}
The request-only strategy evaluates up to \(m_t\) requests for each assignable adversary and therefore requires
\begin{equation}
    O(a_t m_t)
\end{equation}
time. The team-information strategy solves one additional assignment problem between adversaries and requests, with complexity
\begin{equation}
    O\!\left(
        a_t m_t
        \log(a_t\kappa_t)
    \right)
\end{equation}
The full-information strategy additionally simulates the cooperative assignment. It therefore solves a constant number of matching problems of the same form as \eqref{eq:auction_complexity}. This increases computation by a constant number of assignment solves but does not change the asymptotic order.

Over a horizon of \(H\) decision epochs, the cumulative server-side complexities are more precisely expressed as
\begin{equation}
    O\!\left(
        \sum_{t=0}^{H-1}
        n_t m_t
    \right)
    \label{eq:greedy_horizon_complexity}
\end{equation}
for greedy assignment and
\begin{equation}
    O\!\left(
        \sum_{t=0}^{H-1}
        n_t m_t
        \log(n_t\kappa_t)
    \right)
    \label{eq:auction_horizon_complexity}
\end{equation}
for IA and IA-RA.

If \(n_t\leq N\), \(m_t\leq R_{\max}\), and \(\kappa_t\leq\kappa_{\max}\) for all \(t\), these expressions yield the worst-case bounds
\begin{equation}
    O(HNR_{\max})
\end{equation}
for greedy assignment and
\begin{equation}
    O\!\left(
        HNR_{\max}\log(N\kappa_{\max})
    \right)
\end{equation}
for IA and IA-RA.

Table~\ref{tab:complexity} summarizes the per-epoch server-side bounds and the additional cost associated with each adversarial information model.

\begin{table*}[t]
\centering
\caption{Per-epoch computational complexity with precomputed shortest-path
distances}
\label{tab:complexity}
\begin{tabular}{lll}
\toprule
Procedure
&
Primary computation
&
Per-epoch complexity
\\
\midrule
Greedy server assignment
&
Sequential feasible nearest-request matching
&
\(O(n_t m_t)\)
\\

IA server assignment
&
Maximum-cardinality minimum-cost matching
&
\(O(n_t m_t\log(n_t\kappa_t))\)
\\

IA-RA server assignment
&
Repeated maximum-cardinality minimum-cost matching
&
\(O(n_t m_t\log(n_t\kappa_t))\)
\\

Request-only targeting
&
Independent nearest-request search
&
\(O(a_t m_t)\)
\\

Team-information targeting
&
Coordinated adversary--request matching
&
\(O(a_t m_t\log(a_t\kappa_t))\)
\\

Full-information targeting
&
Cooperative simulation and coordinated targeting
&
Constant number of assignment solves
\\
\bottomrule
\end{tabular}
\end{table*}

The analysis shows that incorporating pickup deadlines does not change the asymptotic order of the evaluated assignment procedures when shortest-path distances are precomputed. It adds an \(O(n_t m_t)\) feasibility-screening step, which is of the same order as greedy matching and is dominated by the global matching computation used by IA and IA-RA. The principal computational difference between IA and IA-RA is therefore not their worst-case per-epoch order, but the larger and more frequently reoptimized assignable set under IA-RA.

The next section evaluates how these assignment structures and adversarial information models affect backlog, cancellations, and average cost stability under real-world mobility demand.

\section{Empirical Evaluation}
\label{sec:empirical_results}

We evaluate the proposed framework through the adversarial pickup-and-delivery instantiation developed in Sections~\ref{sec:adversarial_routing} and~\ref{sec:time_feasible_task_allocation}. The experiments are designed to illustrate the operational distinctions established by the theoretical analysis rather than to prove an infinite-horizon stability property from finite simulation.

The evaluation addresses four questions. First, do the selected cooperative fleet sizes and pickup windows produce a meaningful baseline in which persistent failures are not observed? Second, what information is lost when routing performance is assessed through backlog alone and requests never expire? Third, how do assignment structure, adversarial information, and fleet composition affect the accumulation of deadline failures? Finally, does the proposed average cost reveal degradation that is hidden by a backlog-only metric?

\subsection{Simulation Environment and Evaluation Protocol}
\label{subsec:experimental_setup}

We simulate a directed road-network subgraph of San Francisco centered on the financial district with a radius of \(1500\) meters. The resulting graph contains \(1026\) intersections and \(2300\) directed streets. Each edge requires one simulation step to traverse, and each step corresponds to one minute of operation.

Each experimental run contains \(H=5760\) decision epochs, corresponding to four days of simulated operation. All policy--adversary configurations are evaluated using the same sampled request sequences, thereby controlling for differences in realized demand. Results are averaged over \(N_{\mathrm{run}}=100\) independent request-sequence realizations.

For run \(j\), let \(O_t^{(j)}\) and \(F_t^{(j)}\) denote the numbers of outstanding and cumulatively canceled requests at epoch \(t\), respectively. The finite-horizon counterpart of the average policy cost is,
\begin{equation}
    \widehat{J}_{\pi,j}^{\mathrm{avg}}(T)
    =
    \frac{1}{T}
    \sum_{t=1}^{T}
    \left(
        O_t^{(j)}+F_t^{(j)}
    \right)
    \label{eq:empirical_finite_failure_cost}
\end{equation}
The reported policy-cost curves are the sample averages
\begin{equation}
    \overline{J}_{\pi}^{\mathrm{avg}}(T)
    =
    \frac{1}{N_{\mathrm{run}}}
    \sum_{j=1}^{N_{\mathrm{run}}}
    \widehat{J}_{\pi,j}^{\mathrm{avg}}(T)
    \label{eq:average_empirical_finite_failure_cost}
\end{equation}
We also examine the sample-average backlog
\begin{equation}
    \overline{O}_t
    =
    \frac{1}{N_{\mathrm{run}}}
    \sum_{j=1}^{N_{\mathrm{run}}}
    O_t^{(j)}
    \label{eq:average_empirical_backlog}
\end{equation}
and, where relevant, the sample-average cumulative cancellation count
\begin{equation}
    \overline{C}_t
    =
    \frac{1}{N_{\mathrm{run}}}
    \sum_{j=1}^{N_{\mathrm{run}}}
    F_t^{(j)}
    \label{eq:average_empirical_cancellations}
\end{equation}

Because the simulation horizon is finite, a plateau in
\(\overline{J}_{\pi}^{\mathrm{avg}}(T)\) or \(\overline{O}_t\) is interpreted
only as finite-horizon behavior consistent with bounded operation. Likewise,
continued growth is evidence of persistent degradation over the observed
horizon, not by itself a proof of infinite-horizon instability.

\subsection{Demand Model and Cooperative Fleet Calibration}
\label{subsec:estimating_prob_dist}

Following \cite{garces2024approximate}, we estimate the request-generation model from historical San Francisco taxi data \cite{piorkowski2009crawdad}. Pickup and drop-off distributions are obtained from the empirical relative frequencies of trips whose origins and destinations lie within the selected road-network region. The arrival-count distribution \(p_{\eta}\) is estimated from the empirical frequency of the number of requests arriving during each one-minute interval.

Because the empirical arrival-count distribution has finite support, the simulated request process satisfies the bounded-arrival condition in Assumption~\ref{assump:bounded_arrivals}. Experiments with finite pickup windows also satisfy Assumption~\ref{assump:bounded_task_windows}.

We use existing queueing-theoretic fleet-sizing results as a cooperative calibration baseline rather than as an average cost stability guarantee. For IA and IA-RA, let \(d_{\mathcal{G}}(i,j)\) denote shortest-path travel time between nodes \(i,j\in\mathcal{V}\). Let \(\xi\) denote the initial location of a randomly selected cooperative agent, let \(\rho\) and \(\delta\) denote the pickup and drop-off locations of a randomly selected request, and let \(v_{\mathrm{rand}}\) denote the location of an agent immediately after completing a previous trip. The baseline service-time quantity from \cite{garces2024approximate} is,
\begin{equation}
\begin{split}
    D_{\max}
    =
    \max
    \Big\{
        \mathbb{E}
        \left[
            d_{\mathcal{G}}(\xi,\rho)
        \right],
        \mathbb{E}
        \left[
            d_{\mathcal{G}}(v_{\mathrm{rand}},\rho)
        \right]
    \Big\}
    +
    \mathbb{E}
    \left[
        d_{\mathcal{G}}(\rho,\delta)
    \right]
\end{split}
\label{eq:empirical_dmax}
\end{equation}
The corresponding cooperative fleet-size baseline is,
\begin{equation}
    N_{\mathrm{coop}}
    \geq
    \mathbb{E}[\eta]D_{\max}
    \label{eq:cooperative_fleet_baseline}
\end{equation}

The empirical estimates are,
\begin{equation}
    \mathbb{E}[\eta]
    \approx
    1.02,
    \qquad
    \mathbb{E}
    \left[
        d_{\mathcal{G}}(\xi,\rho)
    \right]
    \approx
    17.47
\end{equation}
\begin{equation}
    \mathbb{E}
    \left[
        d_{\mathcal{G}}(v_{\mathrm{rand}},\rho)
    \right]
    \approx
    17.62,
    \qquad
    \mathbb{E}
    \left[
        d_{\mathcal{G}}(\rho,\delta)
    \right]
    \approx
    16.27
\end{equation}
Therefore,
\begin{equation}
    D_{\max}
    \approx
    \max\{17.47,17.62\}+16.27
    =
    33.89
\end{equation}
and
\begin{equation}
    \mathbb{E}[\eta]D_{\max}
    \approx
    34.57
\end{equation}
Rounding upward gives,
\begin{equation}
    |\mathcal{C}|=35
    \label{eq:ia_cooperative_fleet_size}
\end{equation}
for IA and IA-RA.

The greedy policy does not have an analogous fleet-size condition in \cite{garces2024approximate}. We therefore select its cooperative fleet size empirically. We use,
\begin{equation}
    |\mathcal{C}|=50
    \label{eq:greedy_cooperative_fleet_size}
\end{equation}
which produces cooperative finite-horizon behavior comparable to that of IA and IA-RA under the pickup windows used in the subsequent experiments.

Let \(D(\mathcal{G})\) denote the diameter of the road graph. Unless otherwise stated, finite-window experiments use,
\begin{equation}
    w_r
    =
    D(\mathcal{G})
    \qquad
    \text{for every request }r
    \label{eq:default_pickup_window}
\end{equation}
At the time of arrival, any pickup node is within one graph diameter of any agent location. As time passes, the deadline-aware assignment rule in Section~\ref{sec:time_feasible_task_allocation} excludes agent-request pairs that can no longer complete pickup before expiration. The graph-diameter window therefore provides a meaningful baseline that allows feasible service while still converting excessive waiting into cancellation.

\subsection{Evaluated Policy-Adversary Configurations}
\label{subsec:policy_knowledge_configs}
The adversarial experiments consider the Cartesian product of the three assignment structures and three information models introduced in Section~\ref{sec:adversarial_routing}. The assignment structures are greedy, IA, and IA-RA. The adversarial information models are full information, team information, and request only. This produces nine policy-adversary configurations, summarized in Fig.~\ref{fig:policy_knowledge_matrix}.

The configuration matrix is intended to separate two sources of vulnerability. The first is the amount of information and coordination capability available to the adversarial agents. The second is the commitment structure of the assignment policy. Under IA-RA, adversarial reports can repeatedly alter pre-pickup assignments and redirect cooperative agents. Under IA and greedy assignment, captured requests remain blocked until expiration, but repeated reassignment is not permitted. The experiments therefore evaluate the interaction between adversarial knowledge and assignment structure rather than treating either factor in isolation.

\begin{figure*}[t]
\centering
\begin{tikzpicture}[
cell/.style={
rectangle,
draw,
rounded corners,
minimum width=4.2cm,
minimum height=2.1cm,
text width=4.0cm,
align=center,
font=\footnotesize
},
header/.style={
font=\footnotesize\bfseries,
align=center
}
]

\node[header] at (0,3.8) {Omnipotent};
\node[header] at (4.8,3.8) {Team Information};
\node[header] at (9.6,3.8) {Request Only};

\node[header] at (-3.3,1.9) {IA-RA};
\node[header] at (-3.3,-0.2) {IA};
\node[header] at (-3.3,-2.3) {Greedy};

\node[cell] at (0,1.9)
{Strategic coordinated interception of cooperative matching outcomes with reassignment capabilities at every timestep};

\node[cell] at (4.8,1.9)
{Coordinated request blocking guided by request structure and adversarial team composition, with reassignment capabilities at every timestep};

\node[cell] at (9.6,1.9)
{Distributed and uncoordinated adversarial request blocking based only on request locations, with reassignment capabilities at every timestep};

\node[cell] at (0,-0.2)
{Strategic coordinated blocking of cooperative matching outcomes with commitment to matched requests until request expiration};

\node[cell] at (4.8,-0.2)
{Coordinated request blocking guided by request structure and adversarial team information, with commitment to matched requests until expiration};

\node[cell] at (9.6,-0.2)
{Distributed and uncoordinated adversarial request blocking based only on request locations, with commitment to matched requests until expiration};

\node[cell] at (0,-2.3)
{Strategic coordinated blocking of nearest requests to unassigned cooperative agents' locations with commitment to matched requests until expiration};

\node[cell] at (4.8,-2.3)
{Coordinated request blocking guided by request structure and adversarial team composition. Adversaries greedily teleport to the nearest request and remain there until expiration};

\node[cell] at (9.6,-2.3)
{Distributed and uncoordinated adversarial greedy request assignment to closest requests with commitment to matched requests until expiration};

\node[header] at (4.8,4.8) {Adversarial Knowledge Model};
\node[header, rotate=90] at (-4.6,-0.2) {Assignment Policy};

\end{tikzpicture}

\caption{Evaluated policy-adversarial knowledge model configurations. Rows correspond to assignment structures and columns correspond to adversarial information models. IA-RA permits repeated pre-pickup reassignment, whereas IA and greedy do not permit reassignment once requests have been assigned. Increasing  adversarial information enables progressively more coordinated and targeted interception of cooperative assignments.}
\label{fig:policy_knowledge_matrix}

\end{figure*}

\subsection{Cooperative Baseline and Deadline Sensitivity}
\label{subsec:fully_cooperative_time_windows}
We first examine fully cooperative fleets to separate deadline-induced failures from adversarial disruption. IA and IA-RA use \(|\mathcal{C}|=35\), while greedy assignment uses \(|\mathcal{C}|=50\). We consider uniform pickup windows
\begin{equation}
    w_r
    \in
    \left\{
        0.5D(\mathcal{G}),
        D(\mathcal{G}),
        2D(\mathcal{G})
    \right\}
    \label{eq:evaluated_cooperative_windows}
\end{equation}

Fig.~\ref{ref:fully_cooperative_fleet_different_time_windows} shows that when the pickup window is at least one graph diameter, the empirical average cost approaches a plateau over the observed horizon. This behavior indicates that the cumulative cancellation term is not continuing to grow materially in the cooperative baseline and supports the use of these fleet sizes and the graph-diameter window in the adversarial experiments.

When the pickup window is reduced to \(0.5D(\mathcal{G})\), the average cost continues to grow. Requests may then expire even when all agents behave cooperatively because the allowable waiting time is insufficient relative to the spatial scale and current fleet load. This result illustrates that deadline failures need not be adversarial: average cost stability depends jointly on service capacity, demand, assignment decisions, and the operational deadlines imposed by the system
designer.

The simulations do not establish infinite-horizon stability of the cooperative policies. Rather, they identify a finite-horizon operating regime in which persistent deadline failures are not observed and that can serve as a controlled baseline for evaluating adversarial effects.

\begin{figure*}[t]
    \centering
    \includegraphics[height=0.12\textheight]
    {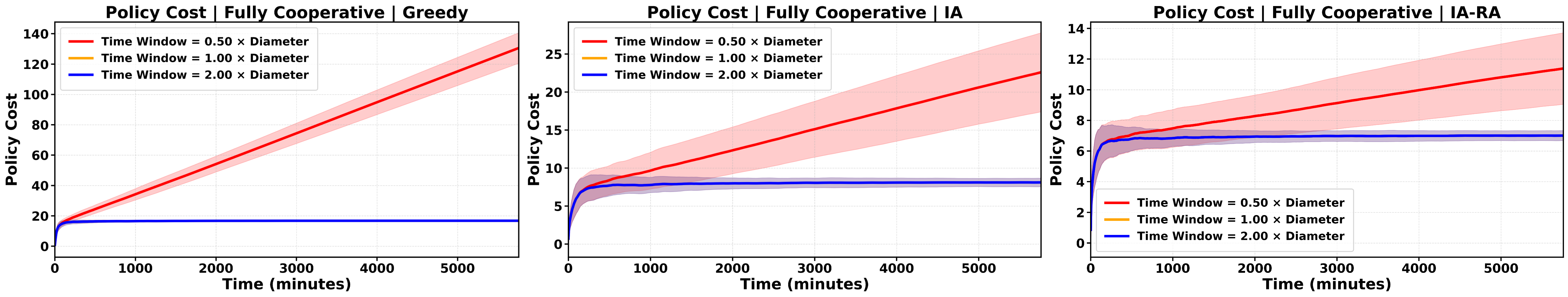}
    \caption{Empirical average cost for fully cooperative fleets under different pickup-window lengths. IA and IA-RA use \( |\mathcal{C}|=35 \), while greedy assignment uses \( |\mathcal{C}|=50 \). When the pickup window is at least one graph diameter, the running finite-horizon cost approaches a plateau, providing behavior consistent with a finite number of deadline failures over the observed horizon. With the shorter window \(0.5D(\mathcal{G})\), cancellations continue to accumulate even without adversarial agents.}
    \label{ref:fully_cooperative_fleet_different_time_windows}
\end{figure*}

\subsection{Backlog-Only Ambiguity Without Request Expiration}
\label{subsec:empirical_no_expiration}

We next remove request expiration from our proposed stage cost and evaluate the adversarial routing system when requests remain outstanding until pickup. In this setting,
\begin{equation}
    |\mathcal{R}_t^{\mathrm{can}}|=0
    \qquad
    \text{for all }t
\end{equation}
and the empirical average policy cost reduces to the running average backlog:
\begin{equation}
    \widehat{J}_{\pi,j}^{\mathrm{avg}}(T)
    =
    \frac{1}{T+1}
    \sum_{t=0}^{T}
    O_t^{(j)}
    \label{eq:no_expiration_backlog_cost}
\end{equation}

Fig.~\ref{fig:degenerate_stability_combined} shows that the backlog eventually approaches a plateau for all evaluated assignment structures, adversarial information models, and fleet compositions. A backlog-only criterion would therefore describe these configurations as bounded over the observed horizon. Nevertheless, many requests remain unserviced for long periods, and the plateau level can be operationally unacceptable.

This experiment illustrates the ambiguity created when delayed requests never expire. Backlog boundedness distinguishes unbounded accumulation from a finite queue, but it does not indicate whether individual requests receive service within a useful time. The result motivates the introduction of operational deadlines: expiration converts excessive waiting into an observable outcome that can be incorporated into the average cost criterion.

Because requests cannot fail in this experiment, it is not itself an
empirical instance of Definition~\ref{def:degenerate_backlog_stability}.
Instead, it demonstrates the underlying limitation of backlog-only
assessment that finite deadlines are intended to resolve.

\begin{figure*}[t]
    \centering
    \vspace{5pt}
    \includegraphics[width=0.99\linewidth]
    {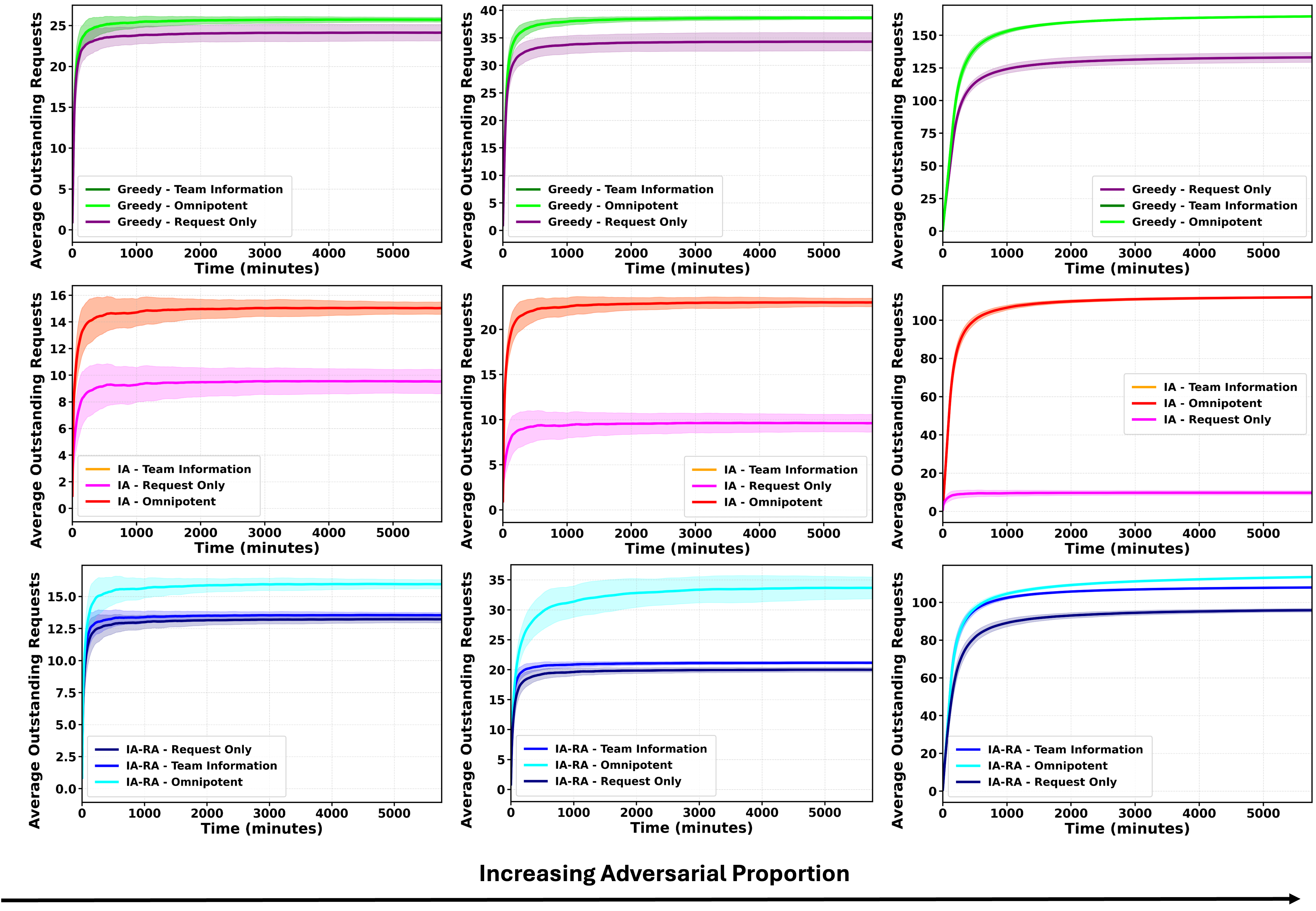}
    \caption{Backlog behavior under adversarial routing without request expiration. Requests remain outstanding until pickup, so the average policy cost reduces to running average backlog. IA and IA-RA use \( |\mathcal{C}|=35 \), while greedy assignment uses \( |\mathcal{C}|=50 \). For IA and IA-RA, the adversarial fleet sizes are $|\mathcal{A}|=7$ $(F\approx0.16)$, $|\mathcal{A}|=15$ $(F=0.3)$, and $|\mathcal{A}|=105$ $(F=0.75)$. For Greedy, the adversarial fleet sizes are $|\mathcal{A}|=9$ $(F\approx0.15)$, $|\mathcal{A}|=22$ $(F\approx 0.31)$, and $|\mathcal{A}|=150$ $(F=0.75)$. Rows correspond to assignment structures and columns correspond to increasing adversarial fleet proportions. Although the backlog approaches a plateau in every evaluated configuration, the plateau may contain a large number of requests that have waited for operationally unacceptable periods.}
    \label{fig:degenerate_stability_combined}
\end{figure*}

\subsection{Persistent Deadline Failures Under Adversarial Routing}
\label{subsec:adversarial_time_window_results}

We now restore finite pickup windows and set
\begin{equation}
    w_r
    =
    D(\mathcal{G})
\end{equation}
for every request. This window produced the cooperative baseline behavior in Section~\ref{subsec:fully_cooperative_time_windows} while remaining finite enough to expose persistent service failures.

Fig.~\ref{ref:policy_costs_combined} reports the empirical average policy cost for the nine policy-adversarial knowledge model configurations across increasing adversarial fleet proportions. Unlike the no-expiration experiment, requests that are repeatedly blocked now become canceled. Their cumulative contribution remains in every subsequent stage cost, so continued cancellations cause \(\overline{J}_{\pi}^{\mathrm{avg}}(T)\) to grow over the observed horizon.

Three patterns are apparent. First, degradation increases as the adversarial fleet proportion \(p_{\mathrm{adv}}\) increases. A larger adversarial proportion can occupy more assignments and prevent a larger portion of the cooperative fleet from servicing requests effectively.

Second, full-information adversaries generally produce the largest average policy cost, followed by team-information and request-only adversaries. Full-information agents can predict which requests would otherwise be allocated to cooperative agents and target those requests directly. Team information enables coordinated blocking but not prediction of the cooperative matching outcome. Request-only agents act without coordination and may redundantly target the same requests.

Third, IA-RA is especially vulnerable to informed adversaries because pre-pickup reassignment permits manipulated reports to repeatedly alter the matching and redirect cooperative agents. IA limits this mechanism by committing agents to selected requests. Greedy assignment also avoids reassignment churn, although it remains vulnerable to persistent local blocking and lacks fleet-level coordination.

These results show that observed finite-failure behavior depends on the interaction between assignment structure and adversarial information. More importantly, the continuing growth of the cumulative-failure cost reveals a form of degradation that is not represented by bounded backlog alone. The observed growth is consistent with violation of average cost stability, although finite simulation cannot establish the corresponding infinite-horizon statement.

\begin{figure*}[t]
    \centering
    \vspace{5pt}
    \includegraphics[width=0.99\linewidth]
    {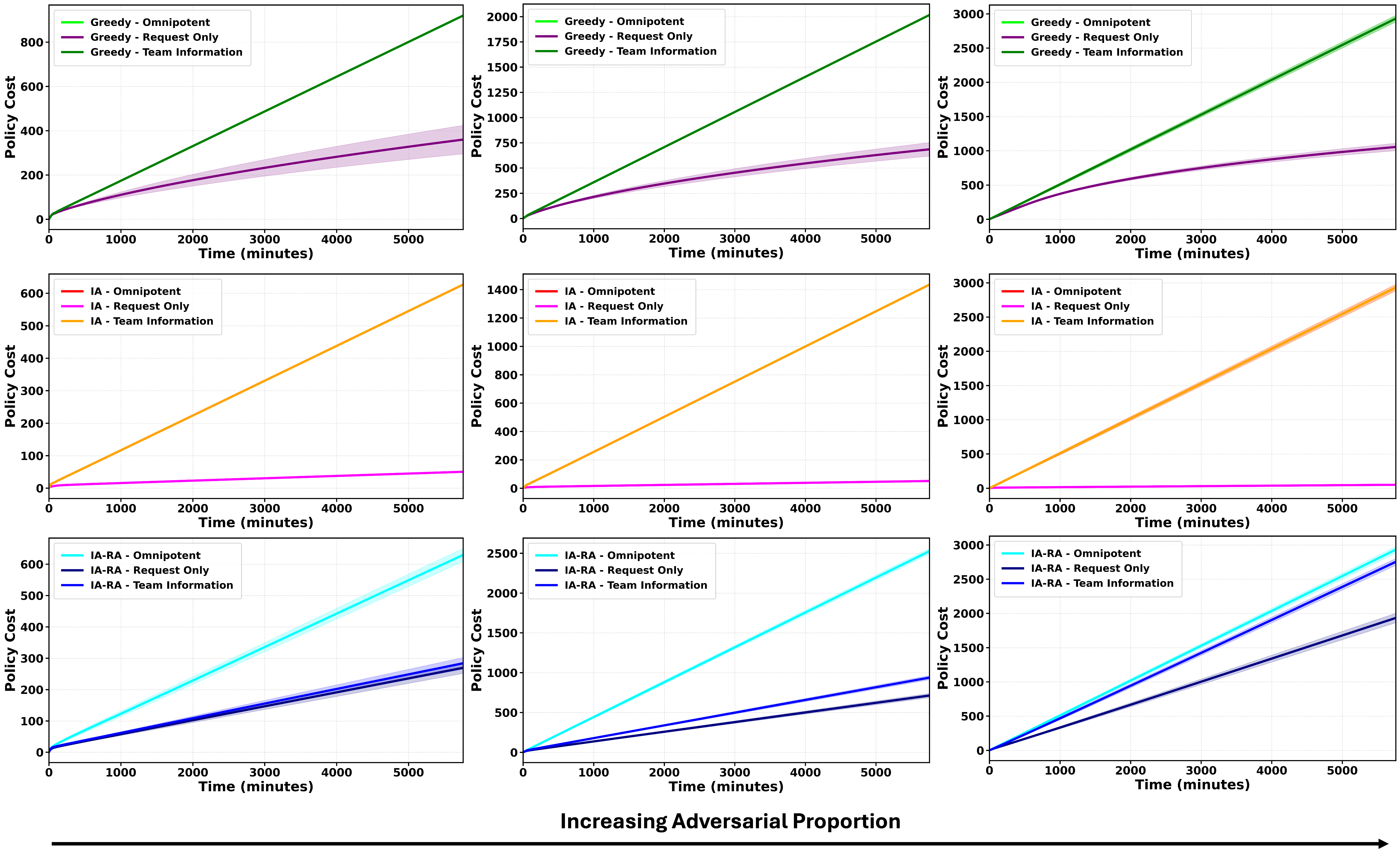}
    \vspace{-10pt}
    \caption{Empirical average policy cost under adversarial routing with pickup windows equal to one graph diameter. Results are shown for greedy, IA, and IA-RA across increasing adversarial fleet fractions and the three information models. IA and IA-RA use \( |\mathcal{C}|=35 \), while greedy assignment uses \( |\mathcal{C}|=50 \). For IA and IA-RA, the adversarial fleet sizes are $|\mathcal{A}|=15$ $(F=0.3)$, $|\mathcal{A}|=35$ $(F=0.5)$, and $|\mathcal{A}|=105$ $(F=0.75)$. For Greedy, the adversarial fleet sizes are $|\mathcal{A}|=22$ $(F\approx 0.31)$, $|\mathcal{A}|=50$ $(F=0.5)$, and $|\mathcal{A}|=150$ $(F=0.75)$. Rows correspond to assignment policies and columns correspond to increasing adversarial proportions. Continued cost growth results from accumulating request cancellations and provides finite-horizon evidence of persistent deadline failure.}
    \label{ref:policy_costs_combined}
\end{figure*}

\subsection{Empirical Signature of Degenerate Backlog Stability}
\label{subsec:comparison_backlog_only}

Finally, we directly compare the proposed average policy cost with a backlog-only metric in the finite-window adversarial setting. This comparison tests the empirical implication of Remark~\ref{rem:bounded_outstanding_requests} and Definition~\ref{def:degenerate_backlog_stability}.

Figure~\ref{ref:policy_costs_comparison_our_policy_cost_and_outstanding_requests} reports both quantities at a representative adversarial fleet proportion \(p_{\mathrm{adv}}\approx0.3\). The backlog-only curves remain bounded because requests leave the outstanding set through either pickup or expiration. Viewed in isolation, these curves would suggest stable operation.

The average policy cost behaves differently. Canceled requests enter the cumulative failure count and remain represented in all subsequent stages. Consequently, the proposed cost continues to grow whenever cancellations persist. The same configurations therefore exhibit bounded backlog and accumulating irrecoverable failures over the observed horizon.

This is the empirical signature of degenerate backlog stability: expiration prevents unbounded backlog growth, but it does not ensure that tasks are successfully serviced. The results support the central distinction of the paper,
\begin{equation}
    \text{Bounded Backlog}
    \;\not\Rightarrow\;
    \text{Average Cost Stability}
\end{equation}

The comparison also clarifies the intended use of the proposed criterion. A backlog metric remains useful for measuring current congestion, while cancellation rate and waiting time quantify the severity of service degradation. The average policy cost provides a complementary reliability test that identifies whether irrecoverable failures continue accumulating rather than disappearing within an acceptable transient.

\begin{figure}[t]
    \centering
    \includegraphics[width=0.99\linewidth]{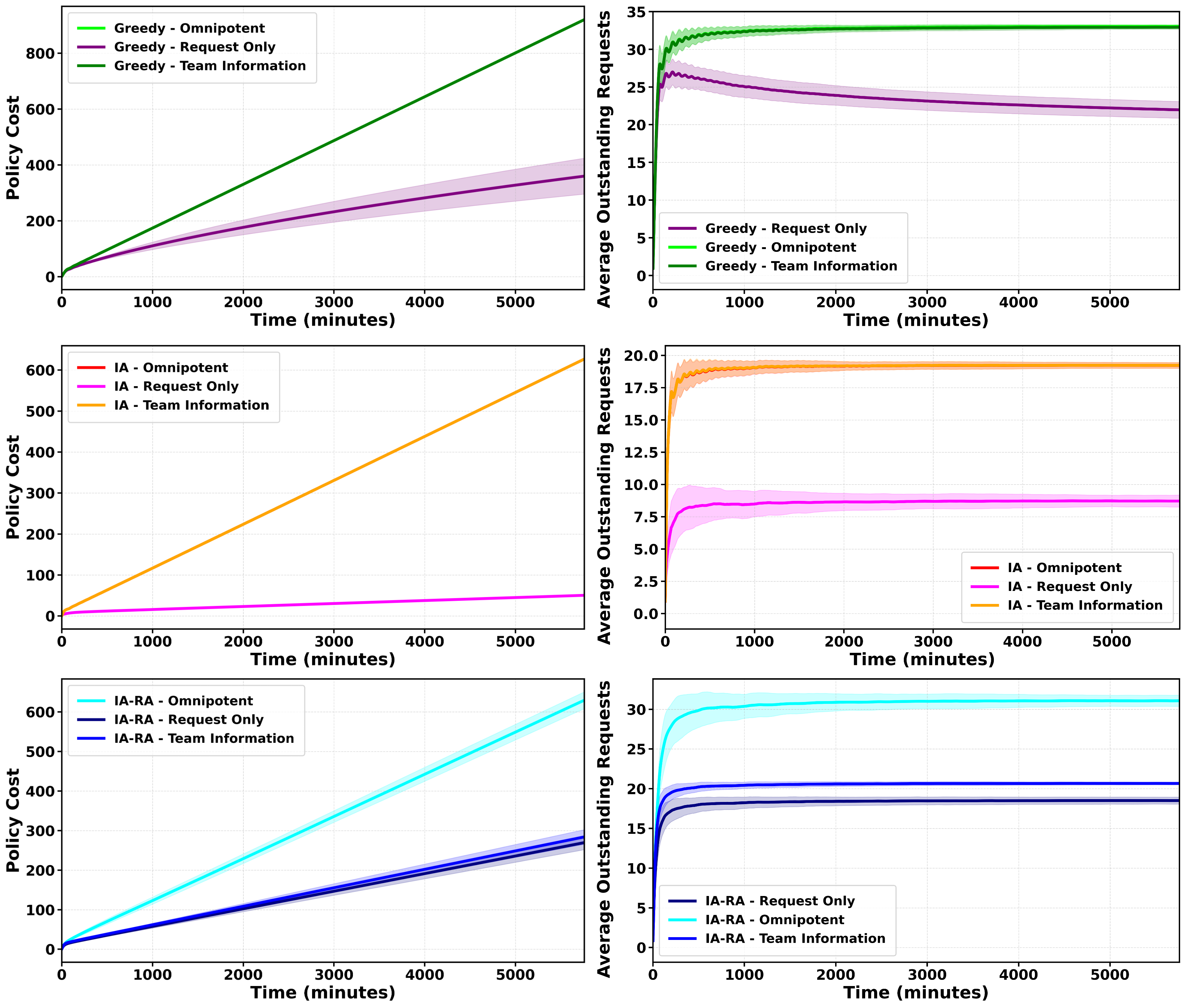}
    \caption{Comparison between the average policy cost and a backlog-only metric under adversarial routing with pickup windows equal to one graph diameter and \(p_{\mathrm{adv}}\approx0.3\). The backlog remains bounded because requests leave the outstanding set through pickup or expiration. The average policy cost continues to grow because canceled requests accumulate. The configurations therefore exhibit the finite-horizon signature of degenerate backlog stability.}
    \label{ref:policy_costs_comparison_our_policy_cost_and_outstanding_requests}
\end{figure}

Taken together, the experiments illustrate the practical implications of the theoretical characterization. Without deadlines, bounded backlog can conceal operationally unacceptable waiting. With finite deadlines, backlog is bounded even when failures persist. Tracking cumulative cancellations therefore reveals whether the system is approaching a failure-free regime or continues to incur irrecoverable service losses under adversarial disruption.

\section{Broader Applicability, Limitations, and Practical Implications}
\label{sec:generalization}

The empirical results in Section~\ref{sec:empirical_results} illustrate the central distinction of the proposed framework: a bounded number of outstanding tasks does not imply that a system has ceased to incur irrecoverable service failures. Although this distinction was evaluated through adversarial pickup-and-delivery routing, it is not specific to transportation or to adversarial behavior. This section identifies the conditions under which the average policy cost framework can be instantiated in other automation systems, discusses its practical use, and clarifies its limitations.

\subsection{Applicability to Other Deadline-Constrained Systems}

The framework applies to automated task-assignment systems satisfying three basic conditions. First, tasks must have identifiable service requirements and finite operational deadlines. Second, a task that misses its deadline must enter a well-defined irrecoverable failure state. Third, the task-management system must be able to observe, or reliably estimate, the numbers of outstanding and irrecoverably failed tasks.

Under these conditions, the application-specific interpretation of the stage cost
\begin{equation}
    g_t
    =
    |\mathcal{R}_t^{\mathrm{out}}|+|\mathcal{R}_t^{\mathrm{fail}}|
    \label{eq:general_application_stage_cost}
\end{equation}
changes, but its mathematical form does not. The quantity \(|\mathcal{R}_t^{\mathrm{out}}|\) represents tasks that remain eligible for service, while \(|\mathcal{R}_t^{\mathrm{fail}}|\) represents the cumulative number of tasks whose required service milestones can no longer be satisfied. The mechanism that causes failure need not be modeled explicitly. Failures may result from stochastic overload, poor assignment decisions, hardware faults, communication loss, sensing errors, environmental disturbances, or non-cooperative agents.

Table~\ref{tab:general_stage_cost} gives representative interpretations in several automation domains. These examples are intended to illustrate how the task states could be defined; they do not constitute empirical validation of the framework in those domains.

\begin{table*}[t]
\centering
\caption{Representative interpretations of outstanding and irrecoverably
failed tasks in deadline-constrained automation systems}
\label{tab:general_stage_cost}
\renewcommand{\arraystretch}{1.2}
\begin{tabular}{p{3.4cm}p{5.2cm}p{6.1cm}}
\toprule
\textbf{Application}
&
\textbf{Outstanding tasks \(|\mathcal{R}_t^{\mathrm{out}}|\)}
&
\textbf{Irrecoverably failed tasks \(|\mathcal{R}_t^{\mathrm{fail}}|\)}
\\
\midrule

Warehouse and fulfillment robotics
&
Orders, replenishment requests, or transport jobs still eligible for
completion
&
Orders or material movements that miss shipping, production, or dispatch
deadlines
\\

Flexible manufacturing systems
&
Production, assembly, or material-handling jobs awaiting completion
&
Jobs that can no longer satisfy production, assembly, or delivery schedules
\\

Hospital logistics
&
Medication, equipment, specimen, or blood-product deliveries still within
their clinical service windows
&
Deliveries that become unusable or clinically invalid after exceeding their
allowable times
\\

Robotic inspection and maintenance
&
Inspection or maintenance tasks whose required completion windows remain open
&
Tasks whose windows expire before the required observation or intervention is
completed
\\

Emergency response and search and rescue
&
Rescue, supply-delivery, or hazard-assessment tasks that remain actionable
&
Opportunities that become infeasible after survivability, access, or response
windows close
\\

Mobility and delivery services
&
Pickup or delivery requests still awaiting service
&
Requests canceled or rendered infeasible after exceeding service deadlines
\\

\bottomrule
\end{tabular}
\end{table*}

The definition of an irrecoverable failure must be chosen carefully for each application. For example, a late warehouse order may still be physically deliverable but may have failed its contractual shipping requirement. A hospital delivery may be irrecoverable if a biological sample is no longer clinically valid, whereas a delayed equipment delivery may remain recoverable. Similarly, an inspection that misses its nominal schedule is not necessarily irrecoverable unless the missed window invalidates the operational objective. The framework therefore requires deadlines and failure states to reflect genuine service commitments rather than arbitrary algorithmic thresholds.

\subsection{Practical Interpretation and Use}
For system operators, the proposed framework suggests monitoring three quantities together:
\begin{equation}
    |\mathcal{R}_t^{\mathrm{out}}|,
    \qquad
    \Delta |\mathcal{R}_t^{\mathrm{fail}}|,
    \qquad
    |\mathcal{R}_t^{\mathrm{fail}}|
    \label{eq:practical_monitoring_quantities}
\end{equation}
where \(|\mathcal{R}_t^{\mathrm{out}}|\) measures current unfinished work, \(\Delta |\mathcal{R}_t^{\mathrm{fail}}|\) measures newly occurring failures, and \(|\mathcal{R}_t^{\mathrm{fail}}|\) records their cumulative effect.

Each quantity answers a different operational question. The backlog \(|\mathcal{R}_t^{\mathrm{out}}|\) indicates current congestion or service pressure. The increment \(\Delta |\mathcal{R}_t^{\mathrm{fail}}|\) measures the immediate failure rate. The cumulative count \(|\mathcal{R}_t^{\mathrm{fail}}|\) reveals whether failures are confined to a transient period or continue to accumulate throughout operation. A bounded backlog should therefore not be interpreted as evidence of reliable operation unless the failure process is also examined.

In practice, average cost stability is most naturally used as a reliability target or diagnostic certificate. For example, an operator could compare the cumulative-failure trajectories observed before and after a dispatch-policy change, a fleet-capacity adjustment, or the deployment of a fault-detection mechanism. If \(|\mathcal{R}_t^{\mathrm{fail}}|\) continues to increase after the expected transient period, the system has not entered a finite-failure regime even if its queue length and average utilization appear acceptable.

The criterion can also support design comparisons. Policies that have similar throughput or average waiting time may differ substantially in whether they continue to produce failures at late operating times. The average policy cost makes this distinction visible, while conventional service-level metrics can still be used to quantify the magnitude and commercial significance of the failures.

\subsection{Limitations of the Proposed Criterion}

Average cost stability is intentionally demanding. It requires the expected total number of irrecoverable failures over the infinite operating horizon to remain finite. For some dynamic stochastic systems, occasional failures may be unavoidable because rare demand surges, hardware faults, or environmental disruptions continue to occur. Such a system may maintain a small and commercially acceptable failure rate while failing the proposed strong criterion.

The criterion should therefore not replace conventional measures such as throughput, average delay, tail latency, cancellation probability, service level, or operating cost. Rather, it distinguishes systems in which failures eventually cease from systems in which failures remain a recurring feature of operation. Whether the strong property is appropriate depends on the application. It is particularly relevant when repeated failures are unacceptable, as may be the case for safety-critical, contractual, or mission-critical tasks.

The theoretical characterization also relies on uniformly bounded arrivals and uniformly bounded task windows. If arbitrarily many tasks can arrive during one epoch, or if tasks may remain serviceable for arbitrarily long periods, the policy-independent backlog bound in Remark~\ref{rem:bounded_outstanding_requests} need not hold. The average policy cost remains well defined, but its equivalence to finite expected total failures may require additional assumptions or a modified analysis.

The framework further assumes that task outcomes can be classified unambiguously as outstanding, serviced, or irrecoverably failed. Applications that permit recovery, reopening, retrying, or partial completion require a more detailed task-state model. A failed attempt should not necessarily be counted as an irrecoverable task failure if the original service objective can still be achieved through a later action.

Finally, the proposed cost is outcome based and does not diagnose the cause of failure. Persistent accumulation of \(|\mathcal{R}_t^{\mathrm{fail}}|\) may reveal insufficient capacity, poor assignments, communication faults, unreliable hardware, or malicious behavior, but additional monitoring or causal analysis is required to distinguish among these mechanisms. Similarly, the present work characterizes stability but does not design a policy that guarantees average cost stability.

\subsection{Scope of the Empirical Evidence and Future Extensions}
The experiments in this paper evaluate only the adversarial pickup-and-delivery instantiation. They demonstrate how the proposed criterion behaves when failures arise from corrupted location reports and unserviced assignments, but they do not establish that the same empirical behavior will occur in warehouse, manufacturing, healthcare, inspection, or emergency response systems. Applying the framework in those domains would require application-specific task definitions, deadlines, failure semantics, dynamics,
and assignment policies.

Several extensions follow naturally. One direction is to derive sufficient conditions under which particular policy classes guarantee average cost stability. Another is to develop finite-horizon statistical tests for determining whether observed cumulative failures are consistent with a transient process or with persistent accumulation. The framework could also be extended to weighted task failures, heterogeneous task criticalities, multiple deadline milestones, or failure costs that depend on the severity of the missed service requirement.

A further direction is to integrate the average policy cost into policy design. The present work uses the criterion for evaluation, but it could also serve as an objective or constraint in adaptive assignment, rollout, reinforcement-learning, or robust optimization methods. In the adversarial routing application, detection and trust-monitoring mechanisms could be combined with the proposed criterion to determine whether removing unreliable agents restores a finite-failure operating regime.

Overall, the proposed framework is best viewed as a general reliability lens for deadline-constrained task assignment. Its principal practical message is that queue length alone is insufficient when tasks can expire: operators must also determine whether irrecoverable failures are confined to a finite transient period or continue accumulating throughout system operation.

\section{Conclusion}
\label{sec:conclusion}

This paper introduces an average cost notion of stability for deadline-constrained automated task-assignment systems. The framework is motivated by a limitation of conventional backlog-based criteria: when tasks have finite service windows, expiration can keep the number of outstanding tasks bounded even while irrecoverable failures continue indefinitely. Similarly, a bounded or even asymptotically zero average failure rate does not imply that failures eventually cease. The proposed criterion addresses this distinction by evaluating both the number of outstanding tasks and the cumulative number of tasks that have irrecoverably failed.

Under uniformly bounded task arrivals and service-window lengths, we show that the number of outstanding tasks is uniformly bounded independently of the assignment policy. Average cost stability is therefore equivalent to a finite expected total number of irrecoverable task failures. This characterization also implies that the total number of failures is finite almost surely. In contrast, policies may satisfy a conventional bounded-backlog condition while continuing to lose tasks, a phenomenon we identified as \emph{degenerate backlog stability}. These results establish that backlog boundedness and average cost stability represent fundamentally different operational properties in systems with task
expiration.

We instantiated the framework in adversarial pickup-and-delivery routing, where internal agents manipulate their reported locations to attract assignments and leave assigned requests unserviced. The instantiation considers multiple assignment structures and adversarial information models and incorporates request deadlines through a feasibility-aware assignment mechanism. This application illustrates why task-level outcomes can provide a more robust basis for reliability assessment when the productive effort of individual agents cannot be directly inferred from their reported states or prescribed routes.

Experiments using real-world San Francisco mobility-on-demand data illustrate the operational consequences of the theoretical distinction. In the absence of request expiration, the backlog approaches a plateau even when many requests experience excessive delays, demonstrating the ambiguity of backlog-only assessment. With finite pickup windows, the backlog remains bounded while cancellations continue to accumulate under adversarial disruption. The average policy cost exposes this persistent degradation, whereas the backlog-only metric does not. The experiments also show that degradation depends jointly on adversarial information and assignment structure, with reassignment-based policies particularly susceptible to repeated manipulation of the matching process. These observations provide finite-horizon evidence consistent with the proposed characterization but do not, by themselves, establish infinite-horizon instability.

The proposed criterion is intentionally strict and should be viewed as a complement to, rather than a replacement for, throughput, waiting-time, failure-rate, service-level, and operating-cost measures. Some systems may tolerate a small recurring failure rate even though they do not satisfy average cost stability. The empirical evaluation focuses on the adversarial routing instantiation. 
Applying the framework in other domains requires application-specific definitions of task deadlines, service milestones, and irrecoverable failure.

Future work will investigate sufficient conditions and policy-design methods for guaranteeing average cost stability, including adaptive assignment, robust optimization, and reinforcement-learning approaches. Additional directions include finite-horizon statistical tests for detecting persistent failure accumulation, extensions to heterogeneous task criticalities and multiple deadline milestones, and integration with trust-monitoring or fault-detection mechanisms. More broadly, the central conclusion of this work is that bounded backlog alone is not a sufficient reliability certificate for deadline-constrained automation. When tasks can expire, reliable operation requires determining not only whether unfinished work remains controlled, but also whether irrecoverable failures eventually
cease.

\bibliographystyle{IEEEtran}
\bibliography{refs}
\vfill

\end{document}